\documentclass[journal,comsoc]{IEEEtran}

\usepackage[T1]{fontenc}
\usepackage{graphicx}
\usepackage{caption}
\usepackage{url}
\usepackage{amsmath,amsthm,amssymb}
\usepackage{mathtools}
\usepackage{mathrsfs}
\usepackage{bm}
\usepackage{color}
\usepackage[utf8]{inputenc}
\usepackage{relsize}
\usepackage{subfig}
\usepackage{float}
\usepackage{booktabs}
\usepackage{adjustbox}
\usepackage{colortbl}
\usepackage{flushend}
\usepackage{cite}

\newcommand{\beq}{\begin{equation}}
\newcommand{\eeq}{\end{equation}}
\newcommand{\beqa}{\begin{eqnarray}}
\newcommand{\eeqa}{\end{eqnarray}}

\makeatletter
\renewcommand*\env@matrix[1][\arraystretch]{%
	\edef\arraystretch{#1}%
	\hskip -\arraycolsep
	\let\@ifnextchar\new@ifnextchar
	\array{*\c@MaxMatrixCols c}}
\makeatother

\begin{document}

\title{Multivariate Time Series characterization and forecasting of VoIP traffic in real mobile networks}

\author{Mario~Di~Mauro,~\IEEEmembership{Senior Member,~IEEE,}
	Giovanni~Galatro, Fabio~Postiglione, \\ Wei~Song, ~\IEEEmembership{Member,~IEEE},
	Antonio~Liotta,~\IEEEmembership{Senior Member,~IEEE}
	 
	\IEEEcompsocitemizethanks{\IEEEcompsocthanksitem M. Di Mauro and F. Postiglione are with the University of Salerno, Italy. E-mails: \{mdimauro,fpostiglione\}@unisa.it
	{\IEEEcompsocthanksitem G. Galatro is with IBM Italy. E-mail: galatro.giovanni@gmail.com}
	{\IEEEcompsocthanksitem W. Song is with the College of Information Technology, Shanghai Ocean University, China. Email: wsong@shou.edu.cn}
	{\IEEEcompsocthanksitem A. Liotta is with the Faculty of Computer Science, Free University of Bozen-Bolzano, Italy. E-mail: Antonio.Liotta@unibz.it}
	\protect\\
		
	}
}

\maketitle
\begin{abstract}
Predicting the behavior of real-time traffic (e.g., VoIP) in mobility scenarios could help the operators to better plan their network infrastructures and to optimize the allocation of resources.
Accordingly, in this work the authors propose a forecasting analysis of crucial QoS/QoE descriptors (some of which neglected in the technical literature) of VoIP traffic in a real mobile environment. The problem is formulated in terms of a multivariate time series analysis. Such a formalization allows to discover and model the temporal relationships among various descriptors and to forecast their behaviors for future periods. 
Techniques such as Vector Autoregressive models and machine learning (deep-based and tree-based) approaches are employed and compared in terms of performance and time complexity, by reframing the multivariate time series problem into a supervised learning one.
Moreover, a series of auxiliary analyses (stationarity, orthogonal impulse responses, etc.) are performed to discover the analytical structure of the time series and to provide deep insights about their relationships.  
The whole theoretical analysis has an experimental counterpart since a set of trials across a real-world LTE-Advanced environment has been performed to collect, post-process and analyze about $600,000$ voice packets, organized per flow and differentiated per codec. 

\end{abstract}


\IEEEpeerreviewmaketitle

\section{Introduction and Motivation}

\IEEEPARstart{P}{erformance} prediction of real-time traffic (such as VoIP) is a crucial topic in the network management field. Predicting and optimizing Quality of Service (QoS) and Quality of Experience (QoE) metrics allows to better dimensioning network infrastructures, improving the battery life of devices, and optimizing the resource allocation strategies \cite{tnsm2,tmc}. This is even more critical in cellular environments, where the high unpredictability of variables such as the interference, but also the concurrency of real-time sessions, and the time-varying load of mobile network nodes pose intriguing challenges. 

We tackle these issues through a multivariate predictive  time series analysis of VoIP traffic across an urban LTE-A environment. At the moment, LTE represents the dominant broadband technology, accounting for $57\%$ of users worldwide  \cite{erireport}. Older technologies such as $2$G and $3$G continue to be intensively used for their robustness, with about $38\%$ of subscriptions; whereas $5$G accounts for about $5\%$ of subscriptions due to its market immaturity. 
Interestingly, one of the most adopted deployment today is the Non-Standalone (NSA) $5$G, where a substantial part of LTE core network is reused to implement voice-based services such as VoLTE \cite{nokiareport}.

The access to LTE technology has stimulated a series of studies devoted to analyzing the performance of QoS/QoE metrics involving, for example: various deployment strategies \cite{lte_deplstrat}, resource allocation \cite{lte_allocat}, probabilistic models \cite{dimaurotnsm}, and coexistence with other technologies \cite{lte_coex}. 


On the other hand, the main contribution offered in this work pertains to a multivariate time series characterization of the dynamic (time-varying) behavior of crucial VoIP metrics which mutually influence each other. Such a cross-dependency has a great impact on forecasting, since the future values of a specific metric (e.g., the bandwidth consumption) will depend not only on the temporal evolution of the same metric, but also on the evolution of other metrics (e.g., round-trip time, jitter) for a given VoIP flow. Accordingly, we formalize analytically such a cross-dependency by means of a vector autoregressive (VAR) model, along with a set of analyses (e.g., stationarity, causality) useful to capture some insights characterizing the mutual influence among the metrics at stake. Such a formalization is then compared to classic (e.g. tree-based) and novel (e.g., deep-based) machine learning approaches. 


As a first step, we carry out an experimental campaign to collect real-world mobile VoIP traffic deriving variables such as bandwidth consumption, mean opinion score (MOS) and signal-to-noise ratio (SNR), among others. In a second step, we perform a predictive analysis aimed at discovering temporal dependencies among the variables and forecast their behavior in future time periods. At this aim, we consider two approaches: $i)$ a statistical approach relying on VAR models, useful to analytically describe the dependencies among interrelated time series; and $ii)$ a machine learning approach, employed by turning a time series structure into a supervised learning problem.
It is worth noting that a time series analysis would be of little use when dealing with data collected in controlled environments (e.g. testbeds). In such a case, in fact, the forecast would be biased since it is possible to manually tune quantities such as interference or noise figures. Conversely, in real settings we deal with uncontrollable variables, which impact the overall performance, such as: time-varying load of radio and network nodes; physical obstacles; weather conditions; and hand over procedures.  

The paper is organized as follows. Section \ref{sec:rw} proposes an overview of similar works, highlighting how our work contributes to the state-of-the-art. In Sect. \ref{sec:netscen}, we offer a description of the experimental environment along with details about the time series construction. In Sect. \ref{sec:models}, we formulate the problem in terms of multivariate time series characterization, and we introduce statistical and machine learning (ML) based models. In Sect. \ref{sec:results}, we present the experimental comparison among the different forecasting techniques by taking into account both performance and times. Section \ref{sec:concl} concludes the work along with some ideas for future research.

\section{Related Work and offered contribution}
\label{sec:rw}

Due to the rapid evolution of telecommunication infrastructures, themes involving the network traffic characterization are becoming decisive from a network management point of view. QoS and QoE metrics, for instance, are typically used as benchmarks to evaluate the quality of a network service; thus, predicting their behavior is crucial to the aim of network optimization and protocol design. Accordingly, in this section we propose an {\em excursus} of relevant works in the field of traffic characterization/forecasting, where we highlight a set of novelties emerging from our work along different directions. 

A first aspect concerns the network traffic forecasting through statistical models, where a common trend is to exploit 
autoregressive moving average (ARMA)~\cite{arma1,arma2} or autoregressive integrated moving average (ARIMA) models~\cite{arima1,arima2,arima3}. Although based on a robust methodology, ARMA and ARIMA models allow to characterize the behavior of individual network variables (in terms of {\em univariate} time series models), but are not able to capture the mutual influence among the variables, which is crucial, for example, to understand the interdependency between objective indicators (e.g., bandwidth) and subjective ones (e.g., MOS).  

A univariate time series perspective is adopted also by that part of the technical literature which employs machine learning models for network traffic forecasting, including neural networks~\cite{ml-forecast1}; support vector machines~\cite{ml-forecast3}; general supervised models~\cite{ml-forecast4}; deep learning models~\cite{ml-forecast2,deep_forecast1,deep_forecast2,deep_forecast3,deep_forecast4,deep_forecast5,deep_forecast6}. 

To fill this gap, we formulate the problem in terms of a {\em multivariate} time series, where each variable is expressed as a function of values of the variable itself and values of the other variables. This approach allows to characterize the interdependency among variables by enabling joint analyses (e.g., orthogonal impulse response analysis) which would have no meaning in a univariate setting. 

Another limitation which emerges in the part of the technical literature focusing on traffic characterization (especially in mobile environments as in our case) is the lack of real-world data. This issue is typically faced through the usage of network simulators, where many variables or models are artificially generated (e.g., mobility models, interference, packet loss, data burst, weather conditions, and many others). 
\noindent Examples include:~\cite{abidi14} and~\cite{carullo16}, where LTE environments are simulated through NS-2; and~\cite{bermudez17,sahu2020,stornig2021}, where some LTE metrics are characterized via NS-3. Other works employ customized LTE simulators to model QoE~\cite{vehic19,cipressi} and QoS indicators~\cite{vienna2,cnsm19}, respectively.  

Even when network experiments are carried out within real mobile scenarios~\cite{real1,real2,real3,real5}, a set of limited metrics are considered, often due to the fact that standard communication protocols (e.g., RTP/RTCP) allow to natively collect only classic metrics, typically relating to bandwidth consumption or network delay. 
To overcome such restrictions we have set up an experimental campaign where, through the RTP Control Protocol Extended Reports (RTCP-XR), we are able to analyze QoS/QoE metrics that are usually neglected in  traffic characterization, including MOS, round trip delay, playout delay buffer, and SNR. 



In summary, the following contributions emerge from our work. First, we formalize the multivariate time series problem of mobile VoIP traffic through the VAR model, which allows to govern analytically the forecasting process. Moreover, through specific analyses including the Dickey-Fuller test, the OLS-CUSUM test and the orthogonal impulse response, we are able to discover interesting insights and hidden relationships among the considered VoIP metrics. 
Then, we turn a set of machine learning techniques (random forest, recurrent networks, etc.) into forecasting methods, by reframing the multivariate time series problem into a supervised learning one through a sliding window approach. This step is needed to evaluate and compare performance and time complexity of the statistical approach against the learning-based ones.
Finally, we remark that the whole time series analysis relies on an experimental campaign carried on a real-world LTE-A network. During this campaign we: $i)$ collect and elaborate a series of VoIP flows exploiting different voice codecs; $ii)$ elaborate a set of performance metrics (most of them neglected in classic literature) through the support of the RTCP-XR protocol.

\section{Network Scenario and Time series construction}
\label{sec:netscen}

\begin{figure}[t!]
	\centering
	\captionsetup{justification=centering}
	\includegraphics[scale=0.44]{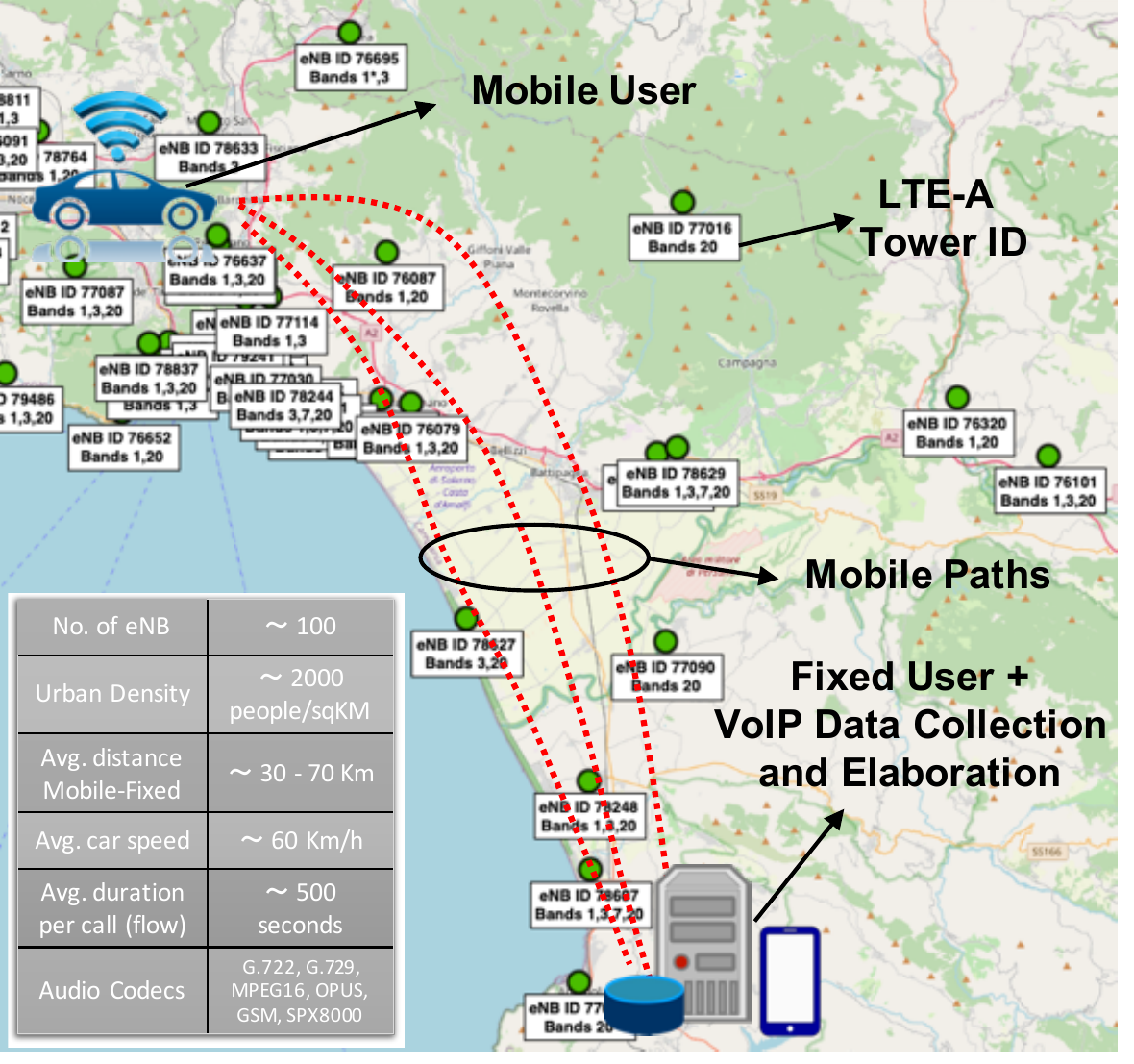}
	\caption{Experimental setting including a mobile user (top-left) and a fixed user with a control server for data collection and elaboration (bottom-right). Main parameters are summarized in the bottom-left table.}
	\label{fig:salerno}
\end{figure}

The location chosen for mobile VoIP traffic collection and analysis is an urban area (about $2000$ people/km$^2$) near Salerno (Italy). Figure \ref{fig:salerno} shows the area map derived from cellmapper \cite{cellmapper}, a crowd-sourced cellular tower and coverage mapping service. The number of evolved nodes B (eNB) aimed at handling radio links amounts approximately to $100$. All the VoIP traffic is collected between two nodes: a mobile node (a car traveling at about $60$ km/h) and a fixed node with a station to collect/elaborate the VoIP flows. The distance between the two nodes ranges from $30$ to $70$ kilometers. Both nodes are equipped with Linphone \cite{linphone}, one of the few softphones supporting the RTCP-XR protocol defined in the RFC $3611$ \cite{rtcpxr}. Such a protocol allows to gather a rich set of metrics not available through the classic RTCP protocol, such as MOS, SNR, round trip delay (or round trip time), and playout delay buffer.  

\begin{figure*}[t!]
  	\centering
  	\captionsetup{justification=centering}
  	\includegraphics[scale=0.45]{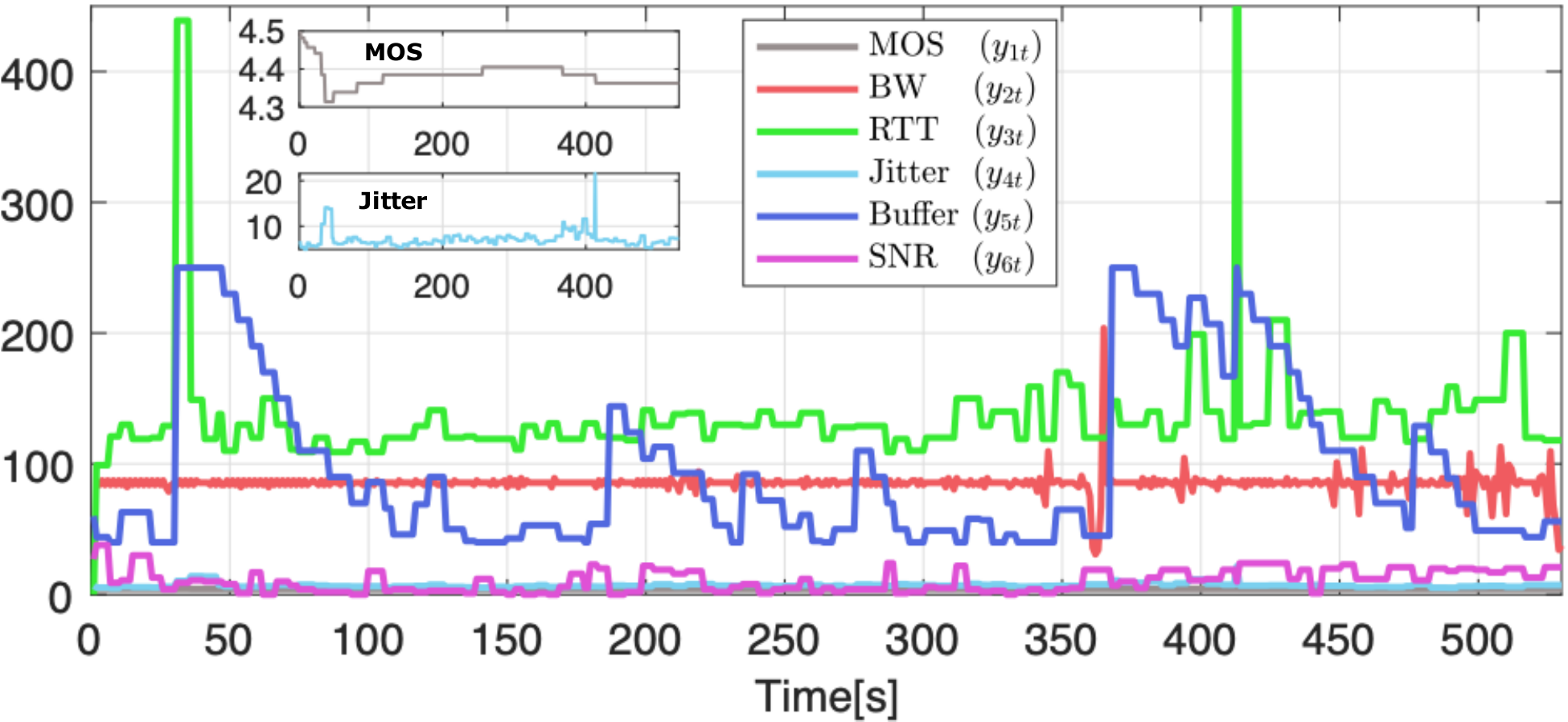}
  	\caption{Time series behavior (Codec G.722  flow) of the six crucial variables: MOS, Bandwidth, RTT, Jitter, Playout Delay Buffer, SNR.}
  	\label{fig:train_test_all}
  \end{figure*}

\begin{figure}[t!]
	\centering
	\captionsetup{justification=centering}
	\includegraphics[scale=0.5]{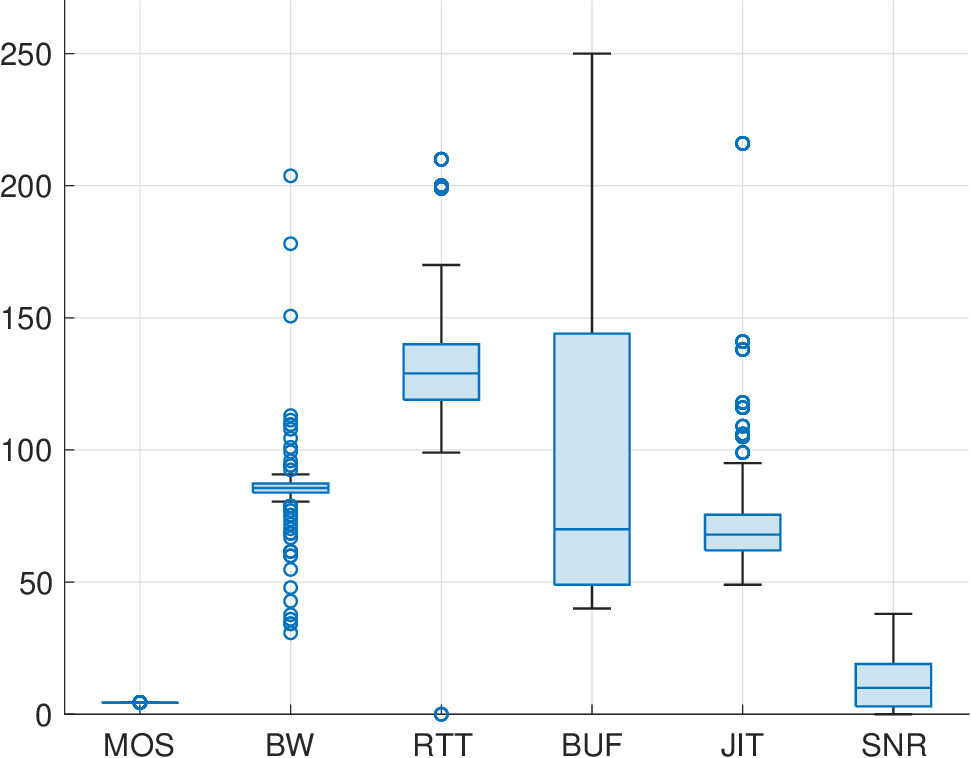}
	\caption{Box-plot representation of time series \\ (Codec G.722 flow).}
	\label{fig:outliers}
\end{figure}

The overall collected flows amount to about 600,000 voice packets, and are divided per codecs, including the following ones: G.722 ($64$ kb/s of bit rate and $16$ KHz of sampling rate); G.729 ($8$ kb/s of bit rate and $8$ KHz of sampling rate); MPEG-16 ($16$ kb/s of bit rate and $16$ KHz of sampling rate); OPUS ($6$ to $128$ kb/s of bit rate and $48$ KHz of sampling rate); GSM ($8$ kb/s of bit rate and $8$ KHz of sampling rate); and SPX-8000 ($8$ kb/s of bit rate and $8$ KHz of sampling rate). Such a choice is justified by the fact that each codec is able to  react differently to diverse network conditions (e.g., the consumed bandwidth or the playout delay buffer) and accordingly adjust the quality of the voice flow. In this way, we obtain a more ample view of the time-based variables behavior and how these are influenced   by the different codecs.
In Table~\ref{tab:stat_param} we report some useful information about the collected dataset including: codec type (first column), number of RTP packets per VoIP conversation (second column), stream length, namely, the duration of conversation (third column), lost packets (fourth column). We derive such information from \textit{RTP stream statistics} section available in Wireshark, the open source sniffer tool aimed at network traffic inspection~\cite{wireshark}.
 \begin{table}[t!]
 	\centering
 	\caption{Some dataset statistics}
 	\resizebox{8.7cm}{!}{
 		\begin{tabular}{cccc}
 			\hline 
 			\textbf{Codec} & \textbf{RTP pkts} & \textbf{Stream length (s)} & \textbf{Lost pkts} \\ \hline
 			G.722 & 55890 & 566   & 0.36\% \\
 			G.729 & 28045 & 447   & 0.4\% \\
 			MPEG-16 & 39181 & 654   & 0.2\% \\
 			OPUS  & 39425 & 405   & 0.4\% \\
 			GSM   & 38357 & 434   & 1.3\% \\
 			Speex-8 & 43128 & 436   & 0.2\% \\
 			\hline 
 		\end{tabular}%
 	}
 	\label{tab:stat_param}%
 \end{table}%
Upon collecting the traffic, we have performed a post-processing stage to extract and process six crucial time-based variables. 
Precisely, for each voice flow (namely for each codec) we built a ($6 \times 1$) time-based vector $y_t=(y_{1t}, \dots, y_{6t})^\mathsmaller{T}$ whose  components are the following six time series:

\begin{itemize}
\item $y_{1t}$: time series representing the {\em MOS}, which quantifies the human subjective experience of a voice call in a dimensionless range between $1$ (low perceived quality) and $5$ (high perceived quality); this metric has been derived from the R-factor (R), a QoE indicator obtainable via RTCP-XR. Then, we have applied the conversion formula provided by ITU-T G.$107$ standard~\cite{itu} to derive the MOS, namely: MOS = $1+0.035 R + 7\cdot 10^{-6} \cdot R(R-60) (100-R) $.
\item $y_{2t}$: time series representing the bandwidth (often {\em BW} for brevity) which provides information about the bandwidth consumption and is measured in kb/s;  
\item $y_{3t}$: time series representing the round-trip time ({\em RTT}), a key performance indicator measuring the time interval (in ms) of a voice packet sent from a source and the ack received from the destination;
\item $y_{4t}$: time series representing the {\em jitter} (measured in ms), namely the variation in voice packet latency evaluated through the formula: $J_n=|({t_r}_{(n)} - {t_t}_{(n)}) - ({t_r}_{(n-1)} - {t_t}_{(n-1)}  )|$ which quantifies the jitter of the $n$-th packet depending on the transmitting (${t_t}_{(n)}$) and on the receiving (${t_r}_{(n)}$) time;
\item $y_{5t}$: time series representing the playout delay buffer (often  {\em Buffer} for brevity and measured in ms), a mechanism to compensate for the encountered jitter by buffering voice packets and playing them out in a steady stream;
\item $y_{6t}$: time series representing the signal-to-noise ratio ({\em SNR}), defined as the ratio between the power of a signal and the power of the background noise, and measured in decibel (dB).
\end{itemize} 
\vspace{10pt}
Figure \ref{fig:train_test_all} shows all the six time series for a single voice flow (G.722 codec). Please note that this representation is meant to offer just a big picture of time series behaviors, since the measurements units are different for each series (e.g., the bandwidth is measured in kb/s, the RTT in ms, the SNR in dB, etc.). At this aim, MOS and jitter have been magnified into two separate insets (MOS in grey and jitter in light blue) to better appreciate their behaviors.
We can preliminarily notice some interesting facts. For instance, it is immediate to see how the RTT time series (in green) has two noteworthy peaks (approximately at $t=30$ and $t=415$ seconds) probably due to some obstacles in the VoIP flow path. Yet, the bandwidth time series (in red) seems to be quite stable (around $80$ kb/s) with a peak at about $t=360$ seconds and a more irregular behavior after $t=450$ seconds, probably due to a more unstable connection. Finally, the jitter time series is more or less regular and lying below $40$ ms, as prescribed by telco standards for VoIP flows \cite{cisco}. 
In order to inspect data variability for this flow, we also report a box-plot representation in Fig.~\ref{fig:outliers}. Such a representation reveals that metrics such a MOS and SNR seem to be quite stable having $8$ and $2$ outliers (onto a stream length of $566$ s, see Table~\ref{tab:stat_param}). The reason is that MOS naturally varies in a bounded range of values, whereas SNR is typically regularized thanks to the underlying codec. Remaining metrics exhibit more instability basically due to the uncontrollable external factors (e.g., interferences, mobility) thus, the number of outliers is greater: BW~($39$), RTT~($36$), Jitter~($52$), Buffer~($97$). 
We finally note that, to add more value to our work we make available: $i)$ raw datasets divided per codec (as described in Table~\ref{tab:stat_param}); $ii)$ post-processed datasets useful to directly test a given forecasting technique~\cite{dataset}.

\section{Problem formulation and Forecasting Models}
\label{sec:models}

In this section we examine in depth the problem of multivariate time series forecasting, by exploiting different techniques. Basically, through such a formalization, we try to predict future values of the time series on the basis of the current information set, namely the set consisting of current and past values of the series. 

Let $y_{1t}, y_{2t}, \dots, y_{Nt}$ be a set of $N$ related variables. The forecast of the $n$-th variable $y_{n,T+H}$ at the end of period {\em T} can be expressed as: 
\beqa
\hat{y}_{n,T+H}&=&f(y_{1,T}, y_{2,T},\dots,y_{N,T},  \nonumber \\  
&& y_{1,T-1}, y_{2,T-1},\dots, y_{N,T-1}, \dots),
\label{eq:forecast}
\eeqa
where $H$ is the forecast horizon (number of future samples to predict), whereas $f(\cdot)$ is a function of the past observations and can be: $i)$ a statistical model, or $ii)$ derived by a machine learning algorithm. One of the most powerful classes of statistical multivariate models for time series forecasting is represented by the vector autoregressive (VAR) models. In contrast, machine learning models exploit their data-driven features to give a prediction of future values of a time series. It is useful to anticipate  that to exploit a VAR model correctly, some preliminary analyses are needed (e.g., stationarity, residual correlation, etc.). Thus, in the next section we formally introduce the VAR model along with its employment into the multivariate time series field.     

We should clarify that the term {\em variable} used in the classic statistics field is equivalent to the term {\em feature} typically encountered in the machine learning realm. To adopt a uniform notation, we will use the term variable  (or time variable), to highlight the temporal dependency.

\subsection{Vector Autoregressive Model}
\label{subsect:var}

The vector autoregressive model is a generalization of the univariate autoregressive (AR) model used for time series prediction. In the classic AR model, a value from a time series is expressed in terms of its preceding values. The number of preceding values chosen for the prediction is called {\em order} or {\em lag}. 
We remark that, in cases where more time series are present, the AR model does not allow to capture their mutual influence. Conversely, in the VAR model a value from a time series can be expressed in terms of preceding values of the time series itself and the preceding values of other time series. It results in a {\em multivariate} time series prediction problem since multiple time series can influence each other. 

Let ${y}_t=(y_{1t},y_{2t},\dots,y_{Nt}) ^\mathsmaller{T}$ be an ($N \times 1$) vector of time series. The $p$-lag vector autoregressive model VAR($p$) has the following form:
\beq
{y}_t = {c} + {\Phi}_1 {y}_{t-1} + {\Phi}_2 {y}_{t-2} + \dots + {\Phi}_p {y}_{t-p} + {\epsilon}_t,
\label{eq:ar}
\eeq
where ${c} = (c_1,\dots,c_N)^\mathsmaller{T}$ denotes an ($N \times 1$) vector of constants, ${\Phi}_k$ an  ($N \times N$) matrix of autoregressive coefficients ($k=1,\dots,p$) estimated by estimating each equation by Ordinary Least Squares (OLS), and ${\epsilon}_t=(\epsilon_{1t},\dots,\epsilon_{Nt}) ^\mathsmaller{T}$ an ($N \times 1$) unobservable zero mean white noise (or residual) vector process with non singular covariance matrix $\Sigma_\epsilon=\mathbb{E}(\epsilon_t \epsilon_t^T)$, being $\mathbb{E}(\cdot)$ the expectation operator.
Let $\phi_{ij}^{(k)}$ be the element of $\mathbf{\Phi}_k$ at row $i$ and column $j$. For instance, a bivariate VAR($2$) model can be expressed in the following matrix form:

\beqa
\begin{bmatrix}[1.5] y_{1t}   \\ y_{2t} \end{bmatrix}
&=&
\begin{bmatrix}[1.5] c_{1}   \\ c_{2} \end{bmatrix} 
+
  \begin{bmatrix}[1.5]
 \phi_{11}^{(1)}  &
  \phi_{21}^{(1)} \\ 
  \phi_{12}^{(1)}  &
   \phi_{22}^{(1)}
  \end{bmatrix}
  \begin{bmatrix}[1.5] y_{1t-1}   \\ y_{2t-1} \end{bmatrix}  \nonumber  \\  
  &+&
    \begin{bmatrix}[1.5]
    \phi_{11}^{(2)}  &
    \phi_{21}^{(2)} \\ 
    \phi_{12}^{(2)}  &
    \phi_{22}^{(2)}
    \end{bmatrix}
     \begin{bmatrix}[1.5] y_{1t-2}   \\ y_{2t-2} \end{bmatrix}
     +
     \begin{bmatrix}[1.5] \epsilon_{1t}   \\ \epsilon_{2t} \end{bmatrix}. 
\eeqa

The preliminary operation to perform when employing a VAR model is to determine the {\em best} lag $p^*$, which allows to build a VAR($p^*$) model embodying most of information of the $N$-dimensional time series. Choosing the optimal lag length is not trivial since many criteria exist (often contradicting each other). We start by applying the Akaike Information Criterion (AIC) \cite{akaike}, a selection rule aimed at minimizing the forecast mean square error (MSE).
Specifically, the approach is to fit VAR($p$) models having orders $p=0,\dots,p_{max}$ and pick the value of $p$ which minimizes the criterion. 

\begin{figure}[t]
	\centering
	\captionsetup{justification=centering}
	\includegraphics[scale=0.5]{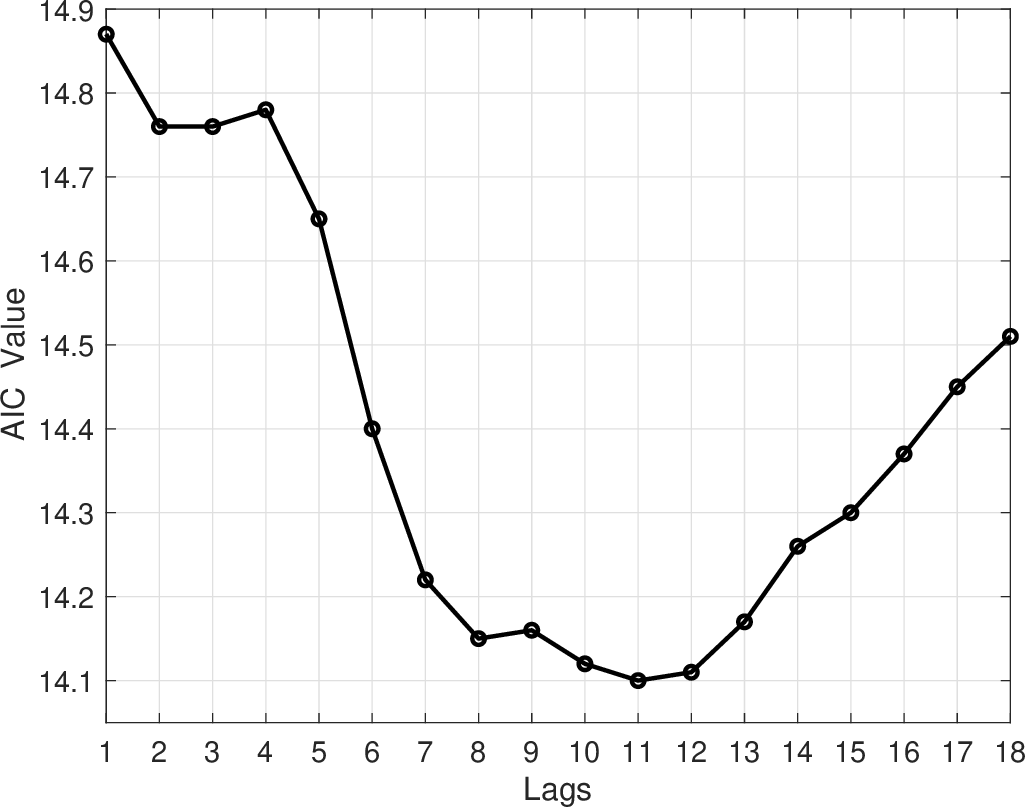}
	\caption{AIC values for different lags applied to the considered multivariate time series model.}
	\label{fig:aic}
\end{figure}

Formally, the AIC criterion obeys to the following rule:

\beq
AIC(p)= \textnormal{log} |\widetilde{\Sigma}_\epsilon| + \frac{2pN^2}{L},
\label{eq:aic}
\eeq
being $L$ the time series length and $|\widetilde{\Sigma}_\epsilon| = L^{-1} \sum_{t=1}^{L} \hat{\epsilon}_t \hat{\epsilon}^T_t$ the determinant of covariance matrix of the residuals estimated by OLS. The results of the AIC criterion applied to our VAR model made of $6$ time series is shown in Fig. \ref{fig:aic}. We can see that the order $p$ which minimizes the AIC criterion amounts to $11$. Actually, when choosing the optimal lag for a VAR model, the lag resulting from a selection criteria (e.g., the AIC) represents only a preliminary choice. 

Often, such a value must be adjusted to encounter also other needs \cite{bank}, such as minimizing the residual correlations as explained in the next subsection.

\subsubsection{Residual Analysis}


When fitting a model to time series data, it is likely to find autocorrelation in the residuals (differences between observed and fitted values). In such a case, the model violates the assumption of no autocorrelation in the errors, and the forecast can be inaccurate \cite{forecast_book}. 
A powerful residuals-autocorrelation test is the Breusch-Godfrey test \cite{godfrey} (also referred to as Lagrange Multiplier, LM test in brief), which considers the VAR model of the error vector  

\beq
\epsilon_t = \Psi_1 \epsilon_{t-1} + \dots + \Psi_h \epsilon_{t-h} + v_t,
\eeq
where $h$ is the maximum lag of the error model and $v_t$ a white noise at time $t$. 
In the LM test, the null hypothesis $\mathcal{H}_0$ is that there is no residual autocorrelation, whereas the alternative hypothesis  $\mathcal{H}_a$ is that residual autocorrelation exists: 

\begin{equation}
\centering
\left\{
\begin{array}{l}
{\begin{array}{ll}
	\hspace{-0.2cm}  \mathcal{H}_0 : \Psi_1= \cdots = \Psi_h=0,
	\end{array}
}
\\
\\
\mathcal{H}_a : \Psi_{\xi} \neq 0 \;\;\; \textnormal{for at least one} \;  \xi \in \{1,\dots,h\}.
\end{array}
\right.
\label{eq:hyp_rescorr}
\end{equation}  

A common way to compute the LM test statistic based on the residuals of the VAR($p$) model is to take into account the following auxiliary regression model \cite{lut2}:

\beq
\hat{\epsilon}_t = c + \Phi_1 y_{t-1} + \dots + \Phi_p y_{t-p} + \Psi_1 \hat{\epsilon}_{t-1} + \dots + \Psi_h \hat{\epsilon}_{t-h} + v_t^*, 
\eeq
where the $\hat{\epsilon}_t$ represent the residuals from the original VAR($p$) model (where $\hat{\epsilon}_t=0$ for $t \leq 0$), and $v_t^*$ is an auxiliary error term. Accordingly, the LM statistic can be computed as

\beq
Q_{LM}(h) = L   \left( N - \textnormal{tr} ( \tilde{\Sigma}_\epsilon^{-1}  \tilde{\Sigma}_v) \right),
\eeq
where $\tilde{\Sigma}_v = \frac{1}{L} \sum_{t=1}^{L} {\hat{v}_t}^*  {\hat{v}_t} ^{*\mathsmaller{T}} $ are the residuals from the estimated auxiliary model and tr($\cdot$) is the trace of a matrix. Under the null hypothesis of
no autocorrelation, it is possible to show \cite{lutk} that $Q_{LM}(h) \xrightarrow{d} \chi^2 (hN^2) $ where $\xrightarrow{d} $ indicates the convergence in distribution (as $L \rightarrow \infty$).  
Moreover, a correction has been proposed by Edgerton and Shukur \cite{es} that exploits the $F$ statistic (based on the Fisher-Snedecor distribution $F(m,l)$ with $m$ and $l$ degrees of freedom) in place of the $\chi^2$, showing interesting results especially in unstable VAR models \cite{hatemi}.
Accordingly, the Edgerton-Shukur (ES) statistic has the following form:
\beq
F(hN^2, \beta),
\eeq
where $\beta=L-N(1+h)+1/2[N(h-1)-1]$.

\begin{table}[htbp]
	\centering
	\caption{Results (in terms of p-values) for models with $p=11$ and $p=12$ lags, respectively. Two residual correlation tests have been considered: Lagrange Multiplier (LM) based on the $\chi^2$ statistic, and Edgerton-Shukur (ES) based on the $F$ statistic, both computed under the null hypothesis.}
	\begin{adjustbox}{max width=\columnwidth}
		\begin{tabular}{ccc|ccc}
			\toprule
			\multicolumn{3}{c|}{\textbf{lag length=11 }} & \multicolumn{3}{c}{\textbf{lag length=12}} \\
			\midrule
			\midrule
			\textbf{$\xi$} & \textbf{p-value (LM)} & \textbf{p-value (ES)} & \textbf{$\xi$} & \textbf{p-value (LM)} & \textbf{p-value (ES)} \\
			\midrule
			1     & \textcolor[rgb]{ 1,  0,  0}{0.0067} & \textcolor[rgb]{ 1,  0,  0}{0.031} & 1     & 0.328 & 0.56 \\
			2     & \textcolor[rgb]{ 1,  0,  0}{0.0044} & \textcolor[rgb]{ 1,  0,  0}{0.036} & 2     & 0.520 & 0.81 \\
			3     & 0.0601 & 0.278 & 3     & 0.604 & 0.90 \\
			4     & 0.0518 & 0.305 & 4     & 0.642 & 0.94 \\
			5     & 0.1336 & 0.554 & 5     & 0.538 & 0.93 \\
			6     & 0.2292 & 0.733 & 6     & 0.244 & 0.78 \\
			7     & 0.1062 & 0.582 & 7     & 0.336 & 0.89 \\
			8     & 0.1414 & 0.691 & 8     & 0.399 & 0.93 \\
			9     & \textcolor[rgb]{ 1,  0,  0}{0.0182} & 0.310 & 9     & 0.125 & 0.75 \\
			10    & \textcolor[rgb]{ 1,  0,  0}{0.0097} & 0.255 & 10    & 0.107 & 0.75 \\
			\bottomrule
			\bottomrule
		\end{tabular}%
	\end{adjustbox}
	\label{tab:res_corr}%
\end{table}%

Once we have chosen the optimal lag value suggested by AIC criterion ($p=11$), we tested the residual correlation through the hypothesis test (\ref{eq:hyp_rescorr}). Such a test has been implemented by exploiting both $\chi^2$ statistic of the LM test and $F$ statistic of the Edgerton-Shukur test. The results are shown in Table \ref{tab:res_corr} - left side - in terms of p-values. Moreover, as suggested by credited literature \cite{lutk,Tsay,forecast_book}, we choose a not so large value for $h$, namely $h = 10$.

By choosing a type $I$ error probability $\alpha=0.05$ to reject the null hypothesis when it is actually true, a p-value lower than $\alpha$ allows us to reject the null hypothesis (no residual correlation), and thus to accept the alternative hypothesis $\mathcal{H}_a$ with an error probability of  $95\%$, at most.
We highlight in red the p-values in correspondence of $\xi$ values where the alternative hypothesis (presence of residual correlation) $\mathcal{H}_a$ of test (\ref{eq:hyp_rescorr}) is accepted, given $\alpha = 0.05$. Such a condition occurs in both LM and ES tests, but the latter seems to be more ``conservative'' and allows  to reject the null hypothesis less frequently than LM test. 
We explore also some higher lags\footnote{Too much high values of lags could lead the system to an overfitting.} and we find interesting results for $p=12$, which represents the second optimal choice from the AIC criterion (see Fig. \ref{fig:aic}). 

The corresponding results in terms of p-values are shown in Table \ref{tab:res_corr} - right side. It is possible to notice that the null hypothesis of no residual correlation is satisfied for all values of $\xi$ in the case of the ES test, with p-values significantly higher than $0.05$. Remarkably, also in the case of the LM test we have p-values higher than $0.05$.   
As mentioned before, we have also tried to further increase the order $p$ of the VAR($p$) model, but we obtained more p-values allowing to reject the $\mathcal{H}_0$ hypothesis of no serial correlation (needed for accurate forecasting) than those obtained for the lag length amounting to $12$, which was finally elected as the optimal choice. 


\subsubsection{Stationarity}
When dealing with VAR models, another important operation consists in removing possible trending behaviors of the involved variables to avoid spurious regressions. Otherwise stated, we have to guarantee the {\em stationarity} of the time series, meaning that first and second moments must be time invariant. 
Pragmatically, the stationarity check is often performed through OLS-based unit root tests. In particular, Dickey and Fuller \cite{dftest} developed a procedure (DF) for testing whether a variable has a unit root or, equivalently, that the variable follows a random walk. 

We use the augmented Dickey-Fuller test (which, differently form classic DF, includes higher-order autoregressive terms in the regression) where the following test model (in its more general form, see \cite{hamilton}) is considered:

\beq
\Delta y_t = \omega_0 + \omega_1 t  +  \theta y_{t-1} + \sum_{k=1}^{p} \delta_k \Delta y_{t-k} + \epsilon_t,
\eeq
where: $\Delta y_t=y_t-y_{t-1}$ is the difference operator, $\omega_0$ is the intercept term (constant), $\omega_1 t$ is the time trend, and $p$ the lag of the autoregressive process. Finally, the test statistic on the $\theta$ coefficient is used to test whether the data need to be differenced to make it stationary. 

The DF test is the following:

\begin{equation}
\centering
\left\{
\begin{array}{l}
{\begin{array}{ll}
	\hspace{-0.2cm}  \mathcal{H}_0 : \theta =0 \;\;\;\;\; (\text{null hypothesis} )   \;\;\;\;\;\;\;\; \text{non-stationarity},  \;\;\;\;\; 
	\end{array}
	}
\\
\\
\mathcal{H}_a : \theta < 0  \;\;\;\;\; (\text{alternative hypothesis} )   \;\;\;\;\; \text{stationarity}.  \;\;\;\;\; 
\end{array}
\right.
\end{equation}  

For our experiments, we have performed the augmented DF test for each variable, verifying that the variables are stationary at first differences, thus, there is no need to apply the differentiation operator. The results are reported in Table \ref{tab:adf_stat}, where the negative values of $\theta$ (second column) for each variable and the corresponding low p-values (third column) suggest to reject the null hypothesis, and to accept the stationarity hypothesis by assuming a type $I$ error probability of 0.05.
\begin{table}[t!]
	\caption {Augmented Dickey-Fuller test per variable.} \label{tab:adf_stat}
	\centering
	\resizebox{.4\textwidth}{!}
	{
		\begin{tabular}{|c|c|c|}
			\hline
			\textbf{Variable}  & \textbf{test statistic} ($\theta$) & \textbf{p-value}  \\
			\hline
			\hline
			\\[-8pt]
			MOS & -3.445 & 9.501 $\cdot$ 10$^{-3}$  \\ \hline
			Bandwidth & -7.974 & 2.718 $\cdot$ 10$^{-12}$  \\ \hline
			RTT & -2.934 & 4.149 $\cdot$ 10$^{-2}$  \\ \hline
			Jitter & -4.157 & 7.780 $\cdot$ 10$^{-4}$  \\ \hline
			Buffer & -2.638 & 8.525 $\cdot$ 10$^{-2}$  \\ \hline
			SNR & -3.192 & 2.044 $\cdot$ 10$^{-2}$  \\ 
			\hline
		\end{tabular}}
	\end{table}

\subsubsection{Stability}
\begin{figure}[t!]
	\centering
	\captionsetup{justification=centering}
	\includegraphics[scale=0.4]{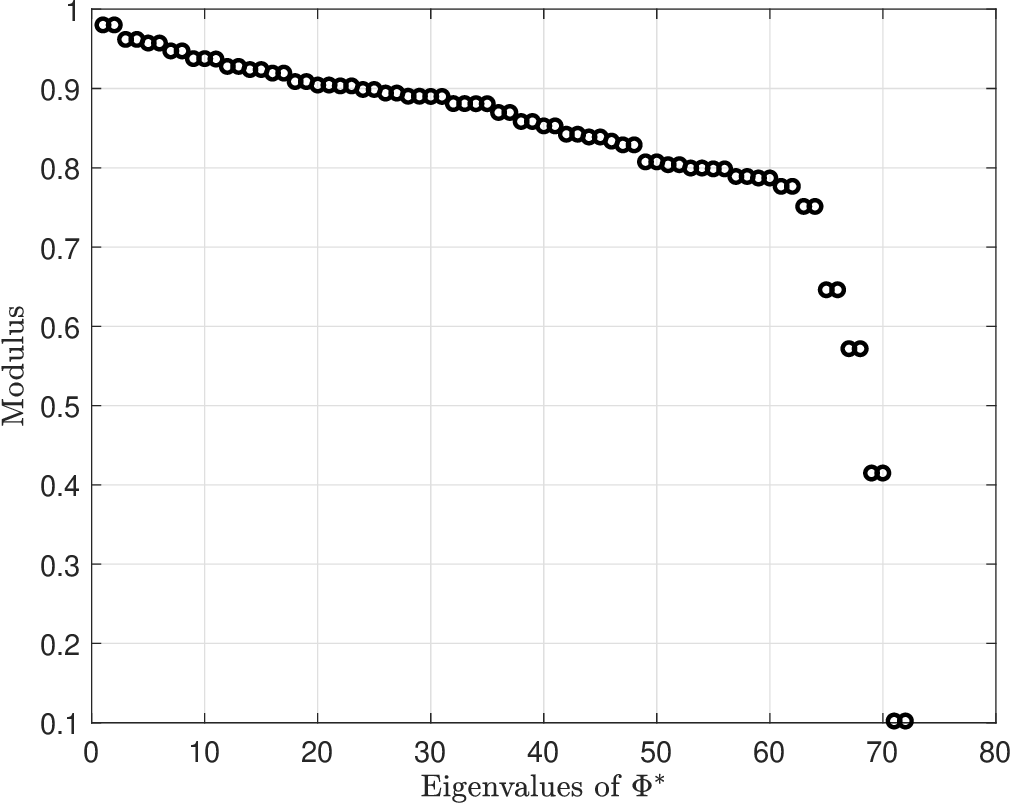}
	\caption{Companion matrix eigenvalues.}
	\label{fig:eigen}
\end{figure}

Stability conditions are typically required to avoid explosive solutions of the stochastic difference equations characterizing a time series expressed in terms of an autoregressive part and a moving average part.
At this aim, it is possible to show \cite{Tsay,lutk} that the VAR($p$) process (\ref{eq:ar}) can be written in the following $Np-$ dimensional VAR($1$) form:

\beqa
\underbrace{\begin{bmatrix}[1.2]   y_t \\  y_{t-1}  \\y_{t-2}  \\ \vdots \\ y_{t-p+1} \end{bmatrix}}_{Y_t}
=
\begin{bmatrix}[1.2] c  \\ 0 \\0 \\ \vdots \\ 0 \end{bmatrix} +
\underbrace{
\begin{bmatrix}[1.2]
	\Phi_{1}  &\Phi_{2}  & \dots  & \Phi_{p-1} & \Phi_p  \\
     I_N & 0  & \dots & 0  & 0   \\
     0 & I_N  & \dots  & 0 & 0 \\
     \vdots & & \ddots & \vdots & \vdots \\
     0 & 0 & \dots & I_N & 0
\end{bmatrix}}_{\bm{\Phi^*}}
Y_{t-1} +
\begin{bmatrix}[1.2] \epsilon_t  \\ 0 \\ 0 \\ \vdots \\ 0 \end{bmatrix}, 
\label{eq:companion}
\eeqa
being $I_N$ the order $N$ identity matrix. 
The process $Y_t$ is stable if the eigenvalues of the {\em companion matrix} $\bm{\Phi^*}$ in (\ref{eq:companion}) have modulus less than one. Such a property is satisfied for the considered VAR($p$) model as can be observed in Fig. \ref{fig:eigen}. Although such an analysis is formally correct to verify the stability condition, it does not allow to capture the behavior in the time domain. Accordingly, we perform in addition an OLS-based cumulative sum (CUSUM)  test \cite{cusum}. Through such a test, it is possible to evaluate the cumulative sums of residuals resulting from the VAR model in order to highlight potential structural changes (a.k.a. structural breaks) in the residuals which can lead to a non-stationary behavior. The test is based on the intuition that if the VAR model coefficients (the autoregressive coefficients) change over the time, the accuracy of the one-step-ahead forecast will decrease and the forecast error will increase. 

\begin{figure}[t!]
	\centering
	\captionsetup{justification=centering}
	\includegraphics[scale=0.5]{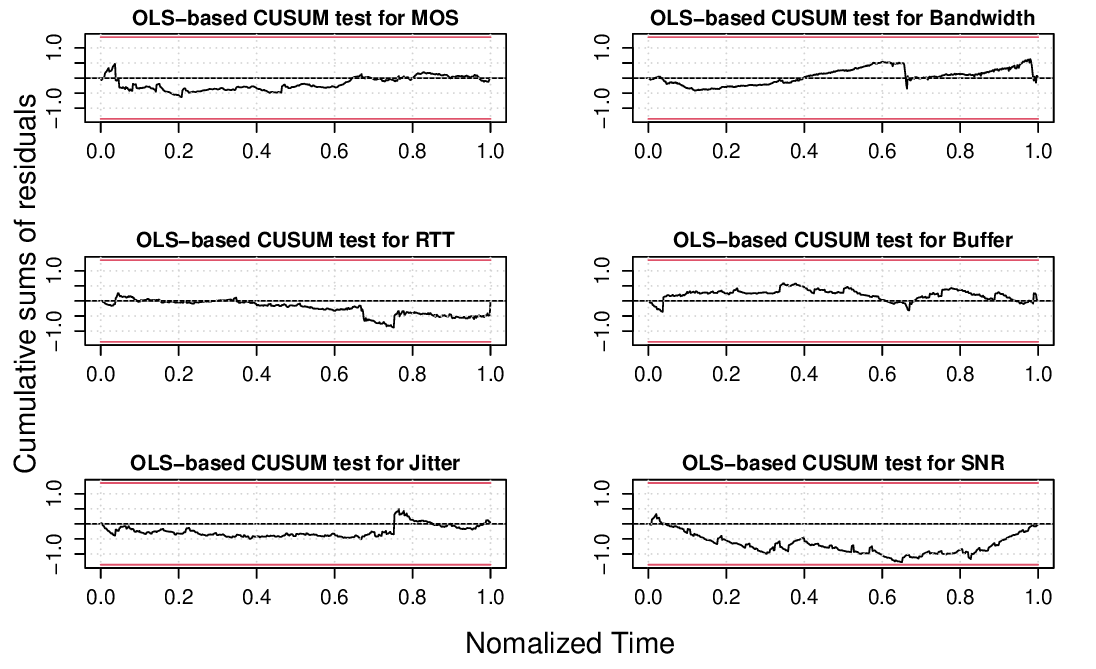}
	\caption{OLS-CUSUM tests applied to the six time series.}
	\label{fig:fluct}
\end{figure}

The panels of Fig. \ref{fig:fluct} show the results of the OLS-based CUSUM test for all the six variables. The x-axis represents the normalized time between $0$ and $1$, where the y-axis reports the cumulative sums of residuals (interpretable as random processes).
It is possible to notice that all the processes are substantially stable with oscillations around the zero value. A slight exception is represented by SNR where is it possible to see some little drifts from the stability value but never exceeding the $95\%$ confidence boundaries (red lines). 

\begin{figure*}[t!]
	\centering
	\captionsetup{justification=centering}
	\includegraphics[scale=0.43]{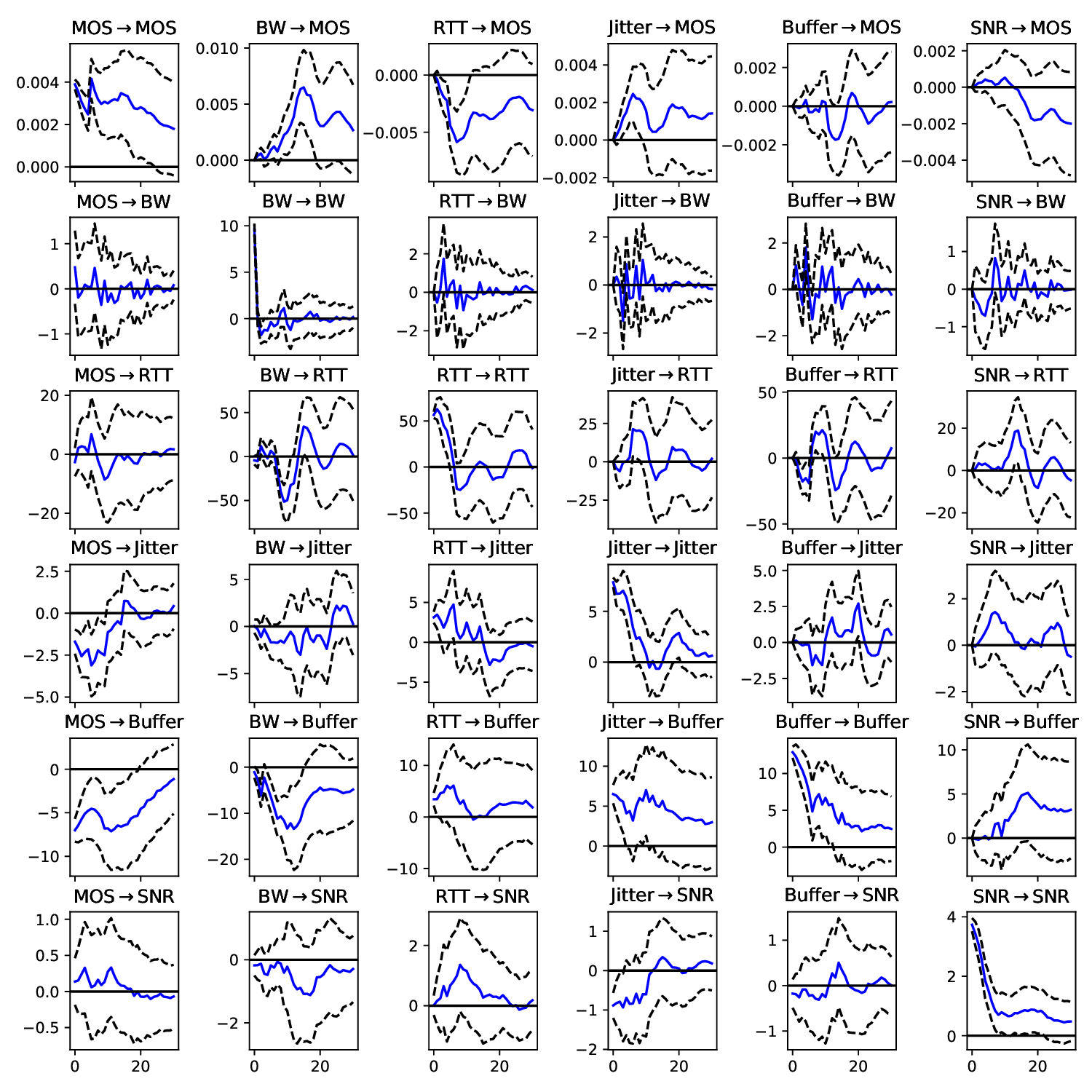}
	\caption{Orthogonal Impulse Response.}
	\label{fig:orth_impulse}
\end{figure*} 

\subsubsection{Time Series relationships}

One of the most interesting aspects when dealing with VAR models is to understand how the time series composing the process are mutually influenced.   
Precisely, it is useful to know the {\em response} of one variable to an external disturbance ({\em unit impulse} or {\em unit shock} in the econometrics jargon) of another variable, which allows to examine more in-depth the cause/effect relation among the involved variables. 
In particular, if we observe a {\em reaction} of one variable to an impulse in another variable, the latter will be {\em causal} for the former \cite{lutk}. In many real-world cases, there is a correlation among the variables in a system. This means that an impulse of one variable should be accompanied by an impulse of some other variables correlated with the modified one. In other words, the impulse response allows to trace the transmission of a single shock within a system of equations \cite{Zivot}. Often, it is interesting to isolate the effect of a single variable shock onto another variable of the system to better capture the interdependencies. At this aim, we implement the orthogonal impulse response functions (OIRF) method \cite{oirf} which allows to rewrite the process (\ref{eq:ar}) as:
\beq
y_t=c+ 	\sum_{i=0}^{\infty} \Phi_i P P^{\mathsmaller{T}} \epsilon_{t-i} = c+ \sum_{i=0}^{\infty} \Theta_i w_{t-i},
\eeq
where: $\Sigma_\epsilon= P P^{\mathsmaller{T}}$ being $P$ a lower triangular nonsingular matrix with positive diagonal elements (also known as {\em Choleski} decomposition, see Appendix A.9.3 in \cite{lutk}), $\Theta_i=\Phi_i P$ and $w_t=P^{-1} \epsilon_t$ being a white noise with covariance matrix $\Sigma_w =\mathbb{E}({w_t w_t^{\mathsmaller{T}}}) = I_K$. Since the white noise errors $w_t$ have uncorrelated components $w_{1t},\dots,w_{Kt}$ with unit variance $I_K$, they are often known as {\em orthogonal} residuals or innovations.	  
Thus, it is reasonable to assume that a change in one component of $w_t$ has no effect on the other components due to orthogonality.
In particular, the $jk$-th element of $\Theta_i$ is assumed to represent the effect on variable $j$ of one unit innovation (namely, one unit standard deviation) in the $k$-th variable that has occurred $i$ periods before.

In our setting, we have $6$ variables resulting in $36$ orthogonal impulse responses\footnote{Historically, such an analysis considers also the effect of an impulse response of a variable on itself.} as shown in the panels of Fig. \ref{fig:orth_impulse}. 
The {\em} causal variables are grouped per columns. 
The x-axis reports the observation period, thus it is possible to evaluate the disturbance effects for various observation periods ($25$ in our case). The blue continuous curves represent the oscillating values of the affected variables around their stability point (horizontal  black line at $0$), namely if the impulse were not applied at all. The black dashed curves surrounding the blue ones represent the asymptotic confidence intervals at 95\% confidence level.
Such an analysis has the merit of highlighting some relationships among variables which are often hidden at a first sight. 
For example, the sub-figure in the first row and second column allows to visualize the effect of a bandwidth shock on the MOS variable (BW $\rightarrow$ MOS). In particular, it is possible to see that a bandwidth impulse causes a slight increase of the MOS by approximately $0.006$ units of innovation after about $15$ observation periods. Then, it decreases. Likewise, a BW impulse causes a decrease of a couple of units of innovation in Jitter after $10$ observation periods before exhausting its effect. Such behaviors are in line with the fact that having more bandwidth is typically beneficial for other metrics.
It is useful to notice that, after a shock, some variables can have a decrease before raising up to their stability point. This is the case of Jitter and Buffer after a MOS impulse which experiment a decrease of $2.5$ and $-5$ units of innovation, respectively (MOS $\rightarrow$ Jitter and MOS $\rightarrow$ Buffer sub-figures). 
Also in this case, we can reasonably admit that a better voice quality can be associated to lower values of jitter which, in turn, is associated to smaller values of the playout delay buffer.   
Interestingly, the mutual influence between two variables can be quite different when the ``causing'' and ``caused'' roles are inverted. For example, in the RTT$\rightarrow$BW case, an RTT shock causes a slight oscillation of BW (with a peak of about $2$ units of innovation around an observation period amounting to $5$) before decaying rapidly to the stability point. In contrast, a BW shock causes a substantial decrease in RTT (BW$\rightarrow$RTT) with two peaks (around $-50$ and $45$ units of innovation) and a slower re-stabilization. Such apparently unusual behavior can be explained by the fact that BW (red curve in Fig. \ref{fig:train_test_all}) exhibits a certain robustness, thus it is not dramatically impaired by unit shocks, whereas RTT (green curve in Fig. \ref{fig:train_test_all}) appears to be more sensitive due to its oscillating behavior, and is then more susceptible to exogenous interventions.   


\subsection{Learning models for time series forecasting}

The application of machine learning techniques to  time series forecasting is a recent issue with interesting applications to econometrics \cite{masini}. When dealing with time series, in fact, the temporal information is crucial, whereas a machine learning dataset is typically a list of information equally treated from a time perspective. This notwithstanding, it is possible to manipulate these models (especially supervised ones) to train on historical time series data and provide future predictions. 
Moreover, some deep learning methods have been explicitly designed to take into account temporal information through memory-based cells, as detailed below.

\vspace{5pt}
\textbf{Recurrent Neural Networks (RNNs)}: such a technique relies on a network architecture able to handle variable-length sequences naturally. In such a way, through the RNNs it is possible to track the state of a system (by retaining past information) and update its state as the time elapses. The memory state is recursively updated with new observations, thus the hidden state $z$ at time $t$ can be represented as a function of the input at time $t$ and the previous hidden state at time $t-1$, namely:
\beq
z(t)=f (z(t-1),y(t)),
\eeq
that, in turn, is used to evaluate the output (namely, the prediction):
\beq
\hat{y}_{t+1} = g(z(t)). 
\eeq
A weak point of RNNs is to manage long-range dependencies connected with transferring information from earlier to later times steps across too long sequences (known as the vanishing gradient problem \cite{bengio}). Such an issue can be solved through the techniques explained below.
The following hyper-parameters have been used for RNN: $30$ RNN units; dropout rate amounting to $0.25$; Adam optimization algorithm (learning rate = $0.1$); tanh activation function; $30$ epochs.

\vspace{5pt}
\textbf{Long Short-Term Memory (LSTM)}: represents an evolved RNN network \cite{lstm} with some internal state cells acting as long-term or short-term memory cells. The output of the LSTM network is modulated by the state of these cells and by three {\em gates} which tune the information flow: the {\em input} gate responsible to update the cell state; the {\em forget} gate in charge of keeping or discarding information on the basis of the input data $y(t)$ and the previous hidden state $z(t-1)$; the {\em output} gate which takes decision about which information to pass to the next LSTM unit. The hidden state at time $t$ is:
\beq
z(t)=o(t) \cdot \textnormal{tanh}(c(t)),
\eeq 
being $o(t)$ the output gate, and $c(t)$ the cell state at time $t$.	 
The following hyper-parameters have been used for LSTM: $30$ LSTM units; dropout rate amounting to $0.25$; Adam optimization algorithm (learning rate = $0.1$); tanh activation function; $30$ epochs.

\vspace{5pt}
\textbf{Gated Recurrent Unit (GRU)}: a lighter version of LSTM \cite{gru} with two gates. The \textit{update} gate which embodies functionalities offered by the LSTM forget and input gates. The \textit{reset} gate which is in charge of deciding how much past information to forget. The GRU hidden state can expressed as:
\beq
z(t)=(1-u(t)) z(t-1) + u(t)  \widetilde{z}(t),
\eeq 
where $u(t)$ is the update gate which decides about the updating amount of its candidate activation $\widetilde{z}(t)$.	
The following hyper-parameters have been used for GRU: $30$ GRU units; dropout rate amounting to $0.25$; Adam optimization algorithm (learning rate = $0.1$); tanh activation function; $30$ epochs.

\vspace{5pt}
\textbf{Convolutional Neural Networks (CNN)}: they are typically used when dealing with classification problems involving spatial information where an image matrix (2D array) is provided to the CNN structure. On the other hand, when dealing with time-series problems, CNNs can be feed with a 1D array since only temporal dimension must be taken into account. Also when applied to temporal data, the CNN uses: $i)$ the convolutional layer aimed at applying filtering to derive the most representative features; $ii)$ the pooling layer to reduce the size of the series while preserving important extracted from convolutional layers; $iii)$ the fully connected layer to map the features extracted by the network into specific classes or values. The following hyper-parameters have been used for CNN: $30$ CNN filters (each of which with size $6$); dropout rate amounting to $0.25$; Adam optimization algorithm (learning rate = $0.1$); tanh activation function; $30$ epochs.

\vspace{5pt}
\textbf{Multi Layer Perceptron (MLP)}: it is the most common form of neural networks and one of the first to be exploited in time series forecasting problems. The lagged observations (say ${x}_i$) are used as inputs of an MLP structure to evaluate the forecast $\hat{y}_{t+1}$:
\beq
\hat{y}_{t+1}=\phi \left( \sum_{i=1}^{n}w_i x_i +b \right),
\label{eq:neuron}
\eeq
where $\phi(\cdot)$ is an activation function (e.g., sigmoid, linear, etc.) to produce the output, and $w_i$ and $b$ are the weights and bias, respectively.
The input data activate the hidden layers (intermediate layers) by following the forward activation direction, and, in turn, hidden layers neurons feed forward into output neurons. 
The MLP process is regulated by the \textit{backpropagation}, a mechanism able to update neurons weights to progressively minimize the error. 
The following hyper-parameters have been used for MLP: $30$ dense units; dropout rate amounting to $0.25$; Adam optimization algorithm (learning rate = $0.1$); tanh activation function; $30$ epochs.

\vspace{5pt}
\textbf{Random Forest (RF)}: a technique based on the {\em bootstrap aggregation} over decision trees. In practice, during the training stage, each {\em tree} within a random forest learns from random samples drawn with replacement ({\em bootstrapping}) so as to reach a lower variance. For each sample $b$, $b=1,\dots,B$, the desired forecast is the average of forecasts of each tree applied to the input data $x_i$, namely
\beq
\hat{y}_{t+1}= \frac{1}{B} \sum_{b=1}^{B} \hat{f}_b(x).
\label{eq:rf}
\eeq
The following hyper-parameters have been used for RF: $30$ estimators (or trees); $10$ as the maximum depth of the tree. 

\textbf{Extreme Gradient Boosting (XGB)}: an improved version of gradient boosting, an iterative technique allowing to fit a decision tree on the residuals obtained from a base learner aimed at improving the prediction model by adding new decision trees. 
The output forecast can be written as:
\beq
\hat{y}_{t+1}=  \sum_{k=1}^{K} \hat{f}_k(x),
\label{eq:gb}
\eeq
where $K$ is the number of trees and $f_k$ is the base learner. An objective function is used to train the model by measuring how well it fits the training data.  
The following main hyper-parameter has been used for XGB in our experiment: $30$ gradient boosted trees.
 
\section{Experimental forecasting results}
\label{sec:results}

In this section we present a comparative analysis of the methods described in the previous section based on the experimental measurements. Before delving into details of the numerical results, we need to provide some clarifications about the processing we have performed on the gathered data. 

\begin{figure}[t!]
 	\centering
 	\captionsetup{justification=centering}
 	\includegraphics[scale=0.26]{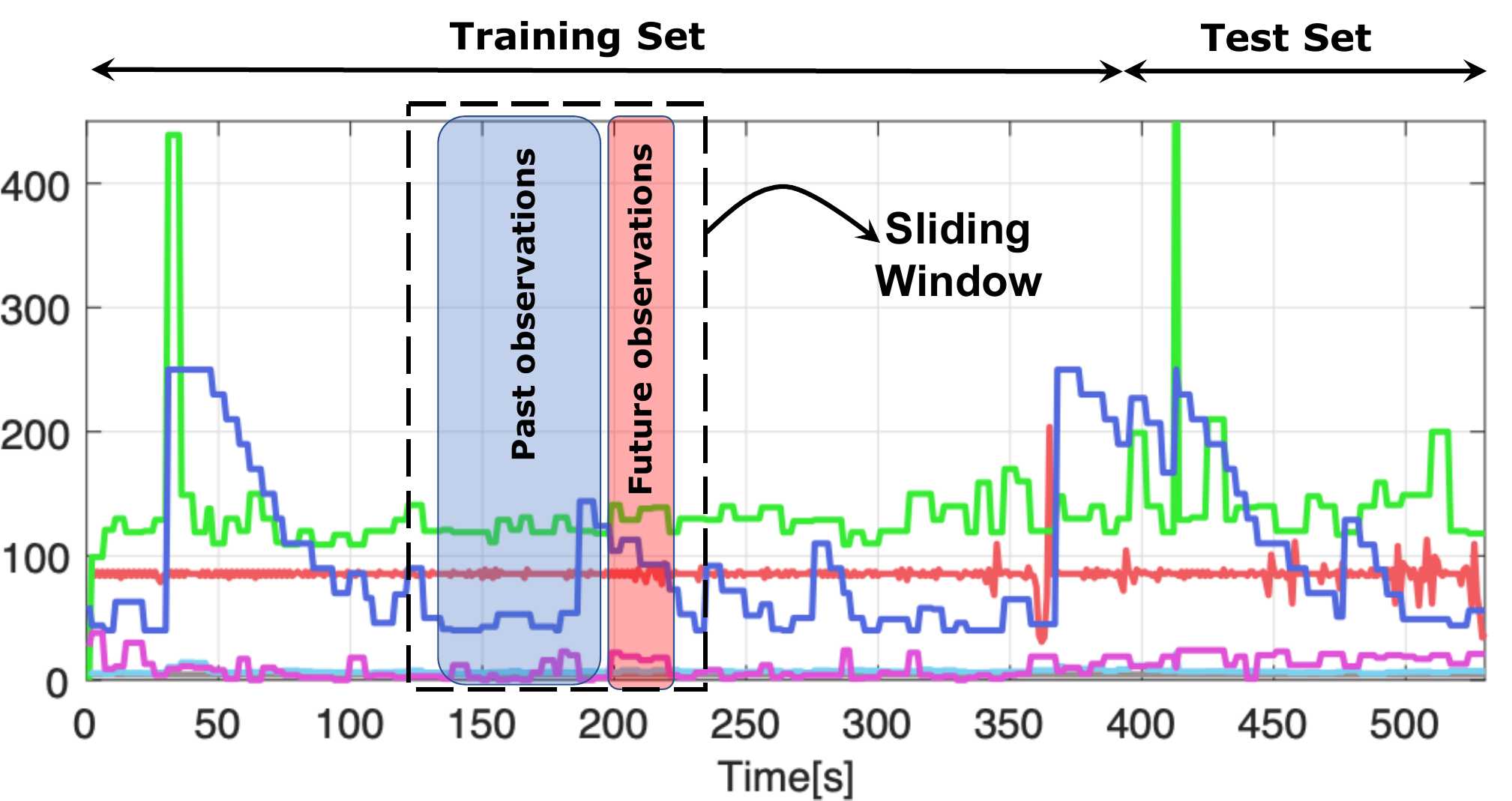}
 	\caption{Time series reframed into supervised learning through the sliding window method. }
 	\label{fig:sliding}
 \end{figure}

A preliminary operation is to re-frame the time series forecasting into a supervised learning problem. We first split  the multivariate time series into training and testing sets, by adopting the classic $70/30$ split ($70\%$ of data is used for training, $30\%$ for testing) as shown in Fig. \ref{fig:sliding}. It is useful to highlight that the classic $k$-fold cross validation method cannot be applied in this setting, which  assumes that there is no relationship among the observations. In contrast, when dealing with time series problems, the temporal {\em continuum} has to be preserved.
Accordingly, we adopt the sliding window mechanism where a part of the input sequence (window of lagged values represented by the past observations within shaded blue area in Fig. \ref{fig:sliding}) is used to forecast new samples (future observations within shaded red area in Fig. \ref{fig:sliding}). The sliding window approach has been profitably employed also in other fields involving time series forecasting such as the smart manufacturing \cite{sliding_manuf} or radar \cite{sliding_radar}. Moreover, as in the aforementioned works, we perform the so-called one-step forecasting where the prediction is made one step at a time to avoid the forecast uncertainty \cite{forecast_book}. 

As regards the tuning of the various learning-based techniques, we have empirically chosen their structures so that the resulting accuracy would be in the same range of the VAR model (which does not require any fine tuning other than the choice of the optimal lag $p^*$). Such an approach is in line to what suggested by credited literature \cite{bengio_book}. 
\begin{figure*}[t!] 
	\centering
	\begin{tabular}{cccc}
		\subfloat{\includegraphics[scale=0.33]{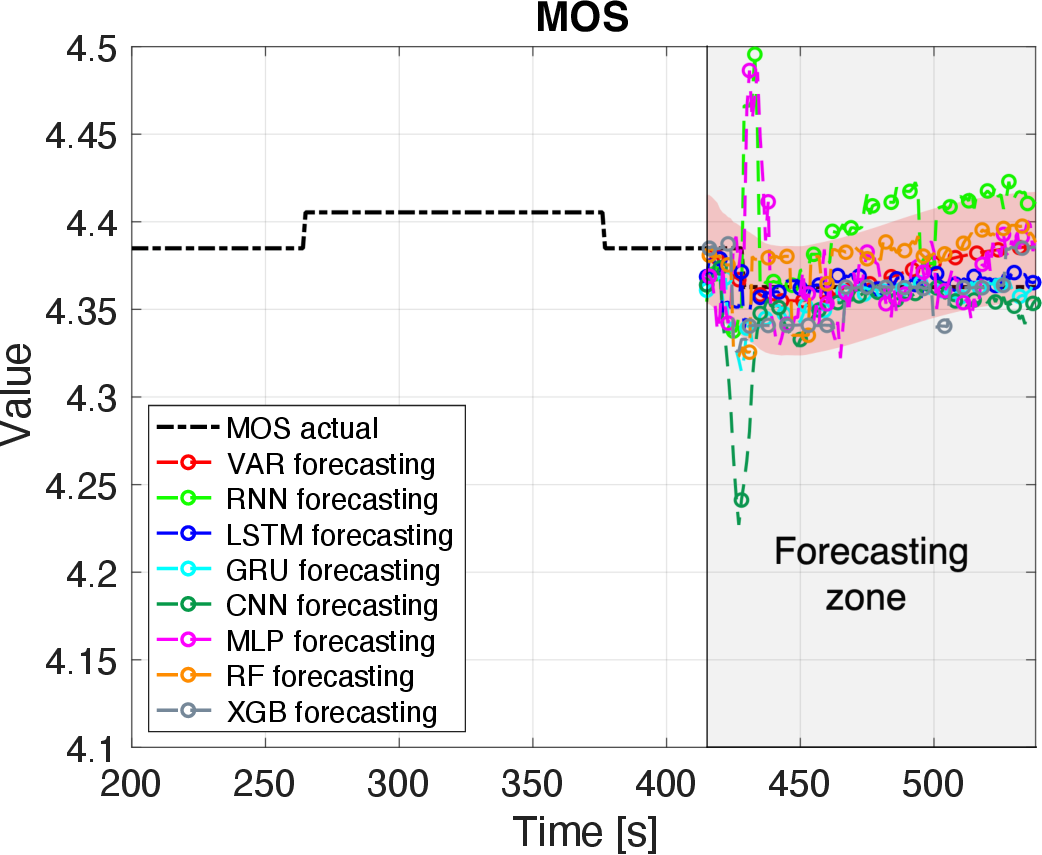}}  \hspace{2mm}
		\subfloat{\includegraphics[scale=0.33]{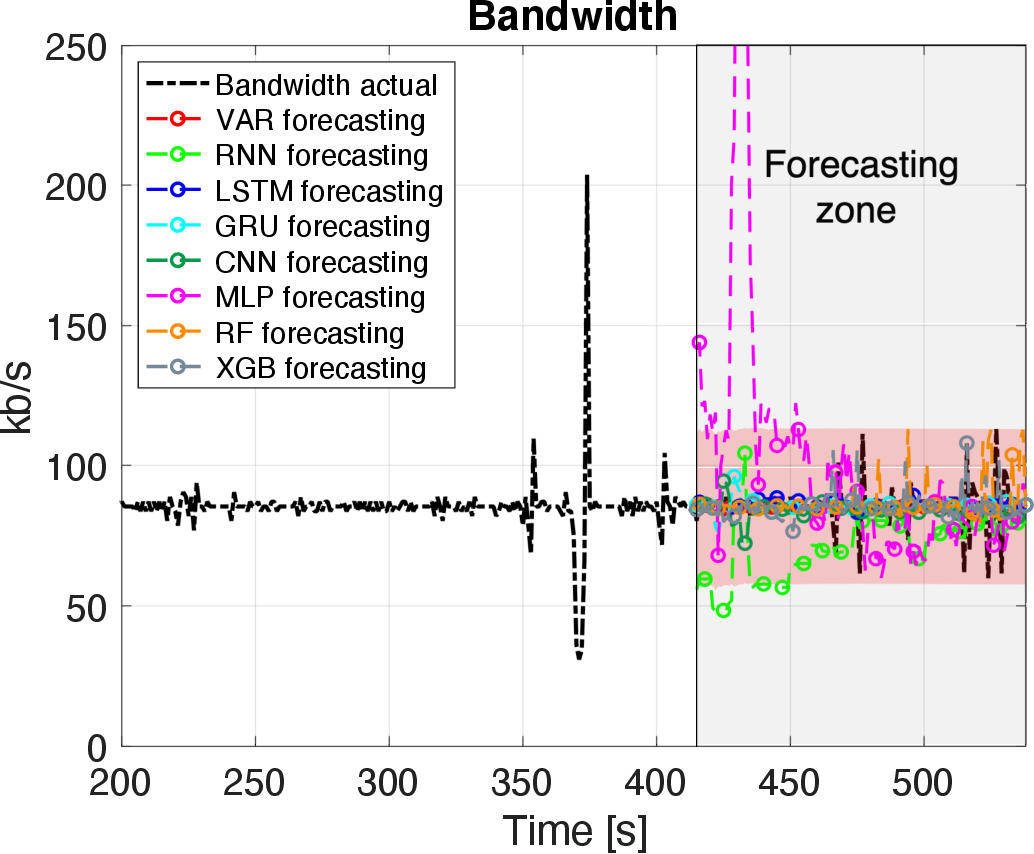}}  \hspace{2mm}
		\subfloat{\includegraphics[scale=0.33]{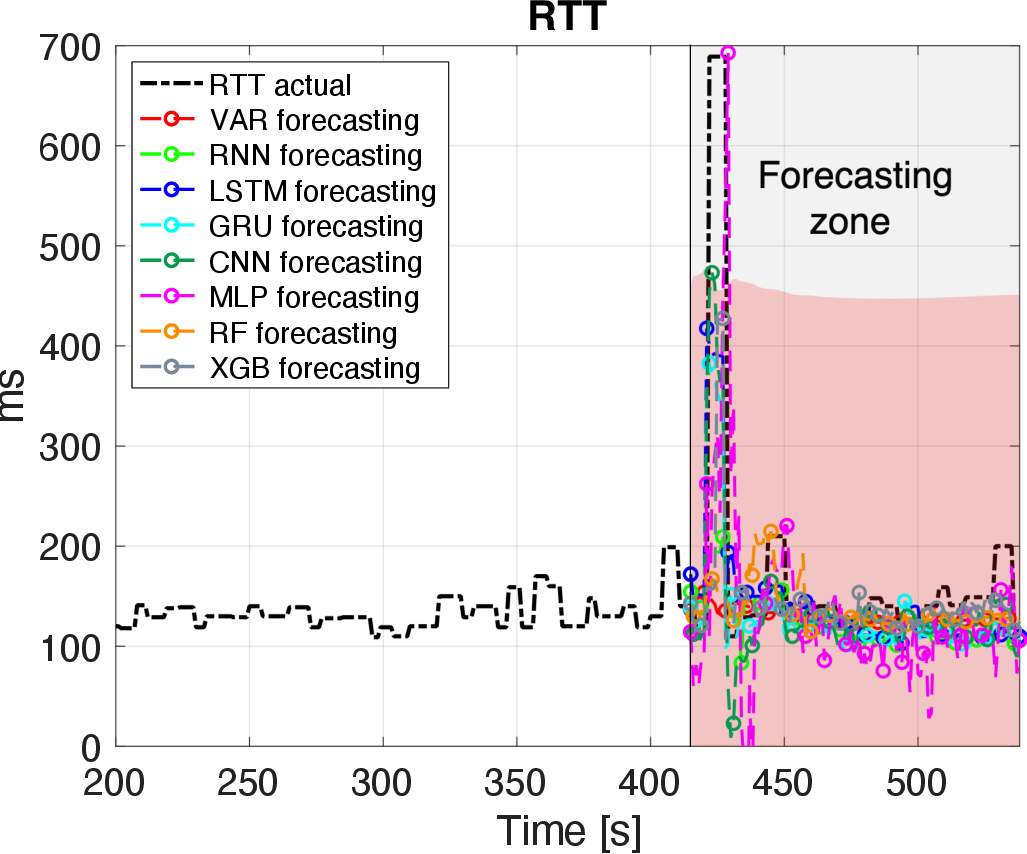}} \\
		\subfloat{\includegraphics[scale=0.33]{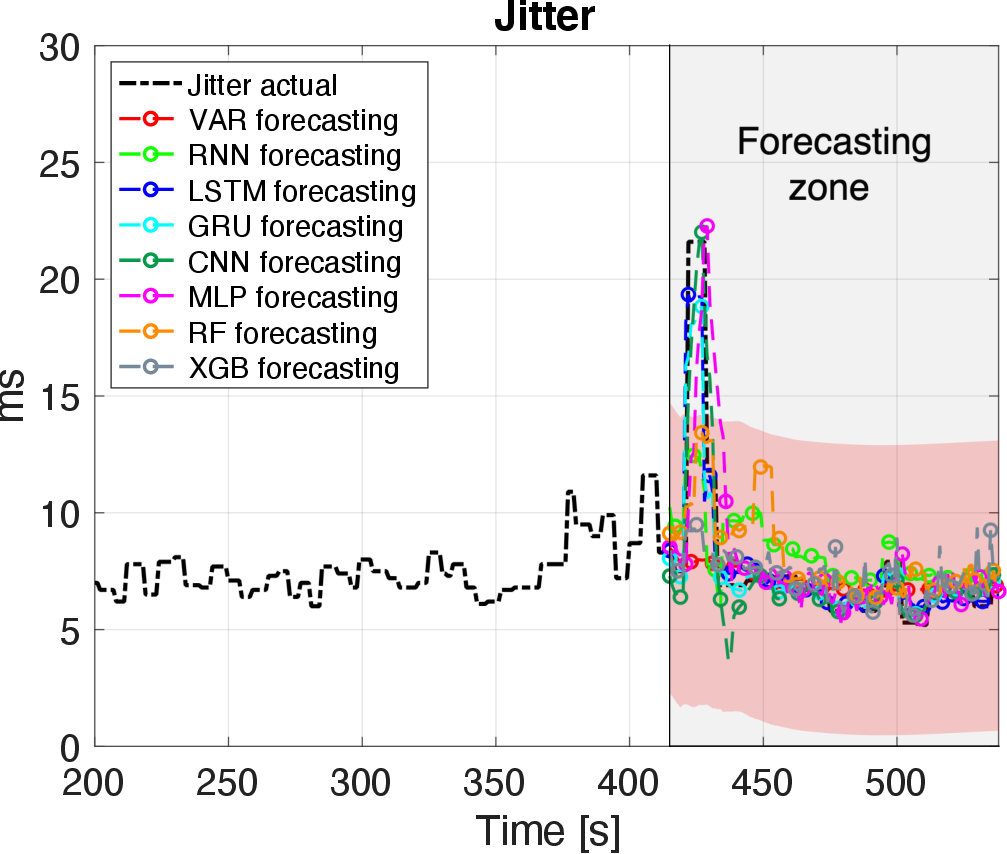}}  \hspace{2mm}
		\subfloat{\includegraphics[scale=0.33]{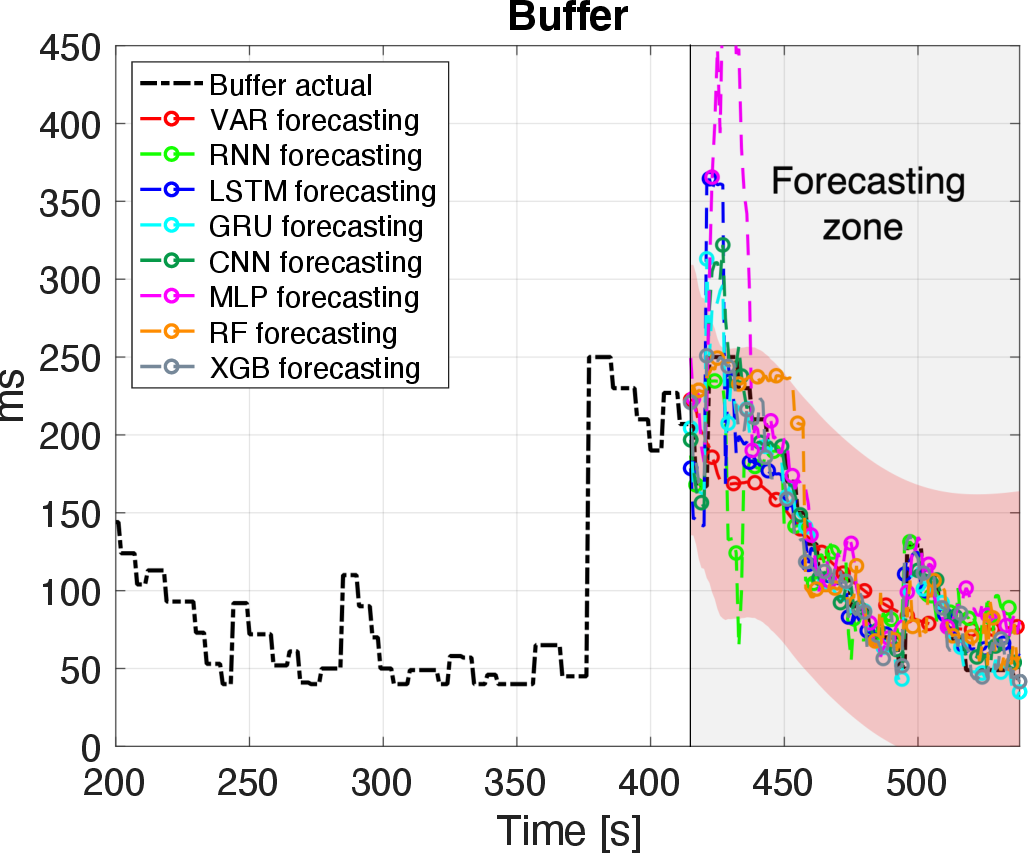}} \hspace{2mm}
		\subfloat{\includegraphics[scale=0.33]{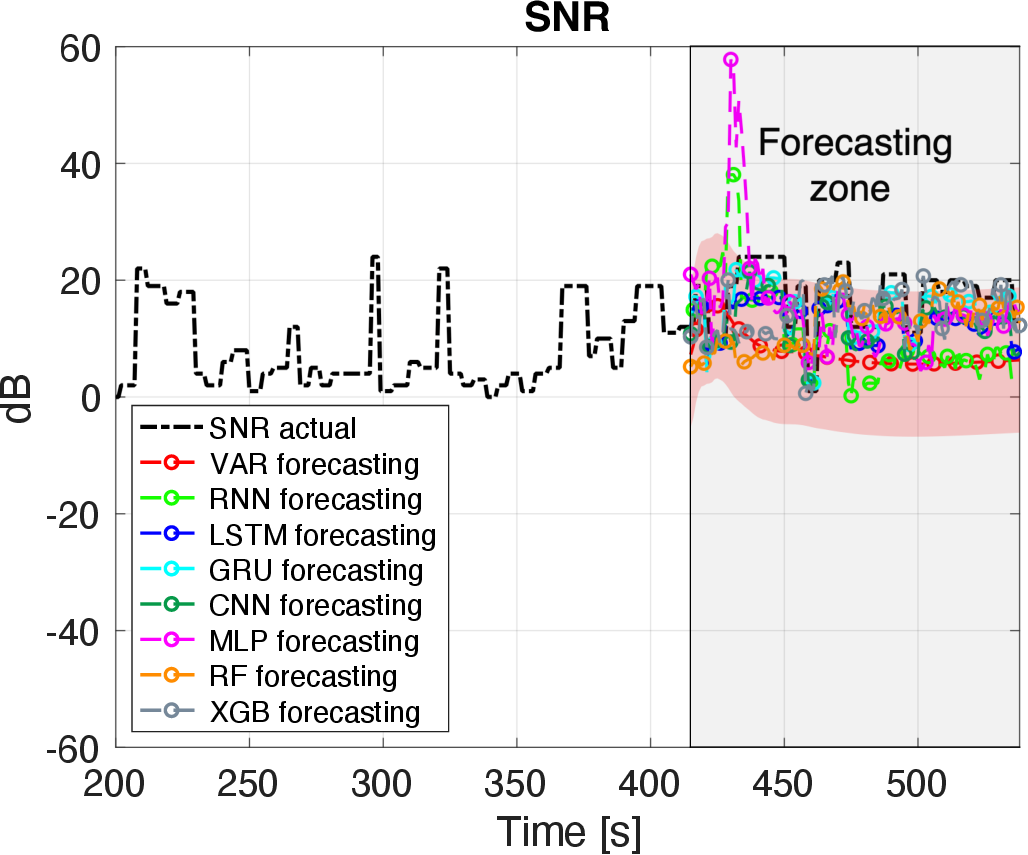}} 
	\end{tabular}
	\caption{G.722 codec: Multivariate time series forecasting for: MOS, Bandwidth, RTT, Jitter, Buffer, SNR, along with the 95\% forecasting intervals in shaded pale red. In the gray area on the right, the results of forecasting for each technique.}
	\label{fig:forecast}
\end{figure*}

\begin{figure*}[t!] 
	\centering
	\begin{tabular}{cccc}
		\subfloat{\includegraphics[scale=0.33]{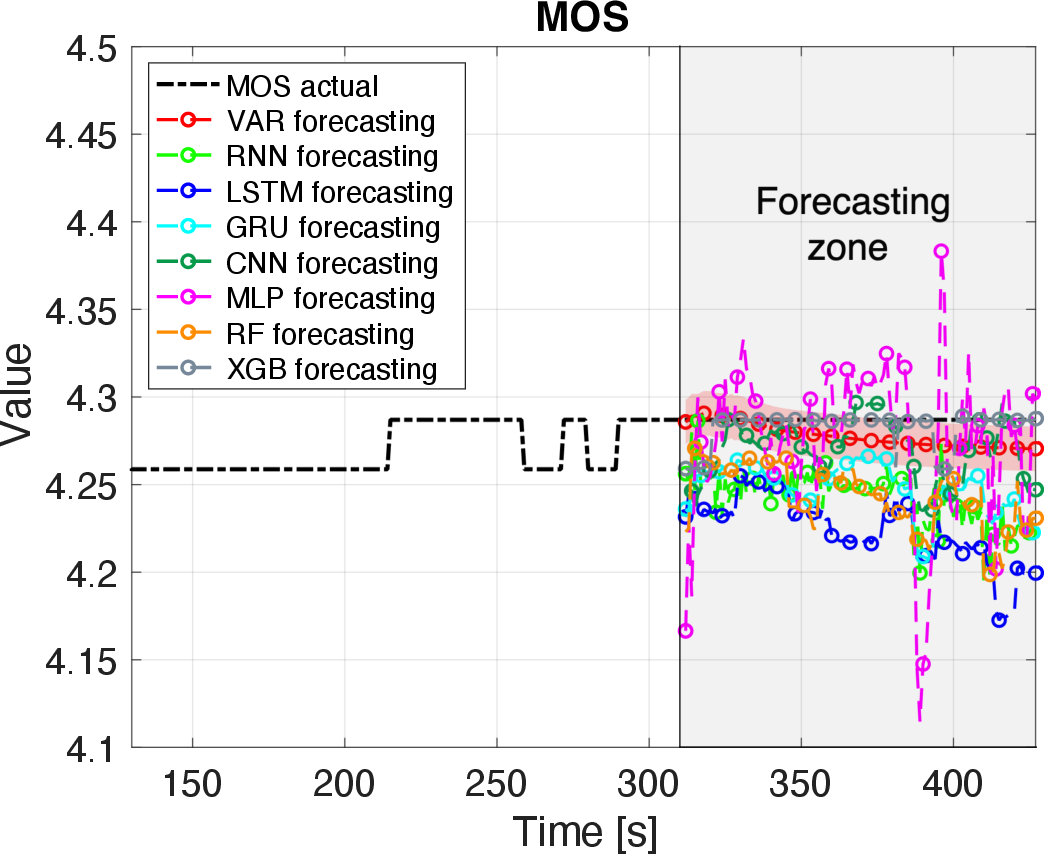}}  \hspace{2mm}
		\subfloat{\includegraphics[scale=0.33]{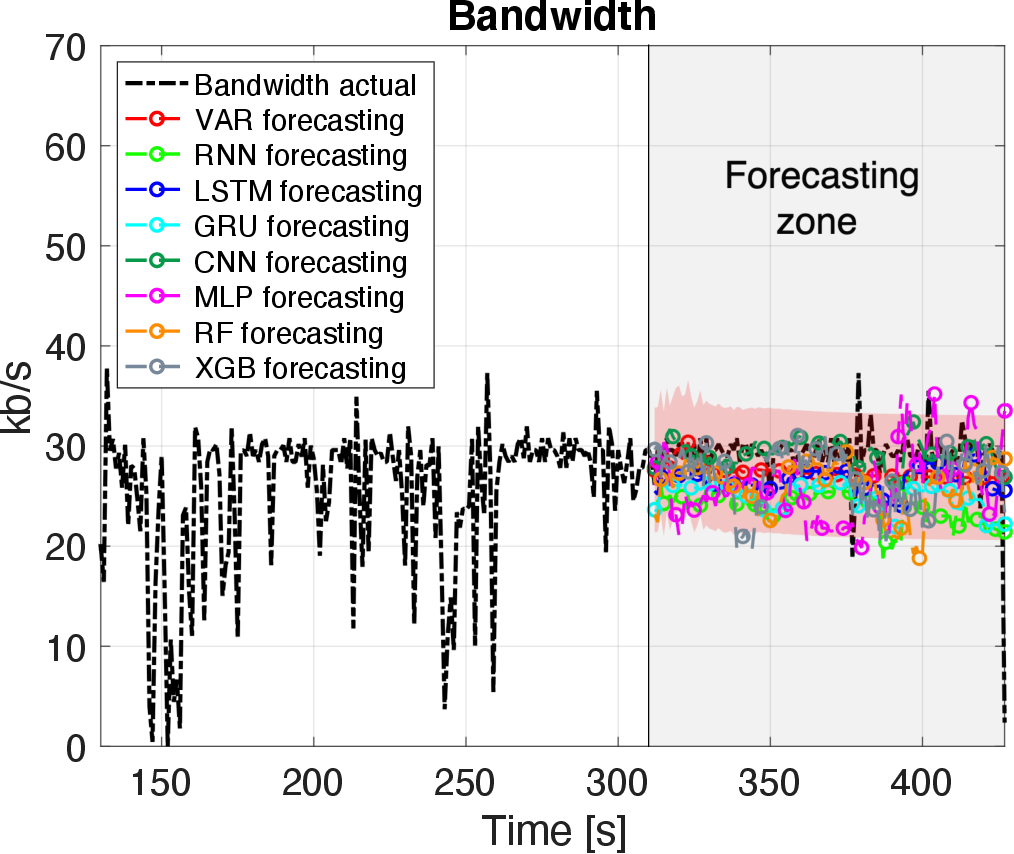}}  \hspace{2mm}
		\subfloat{\includegraphics[scale=0.33]{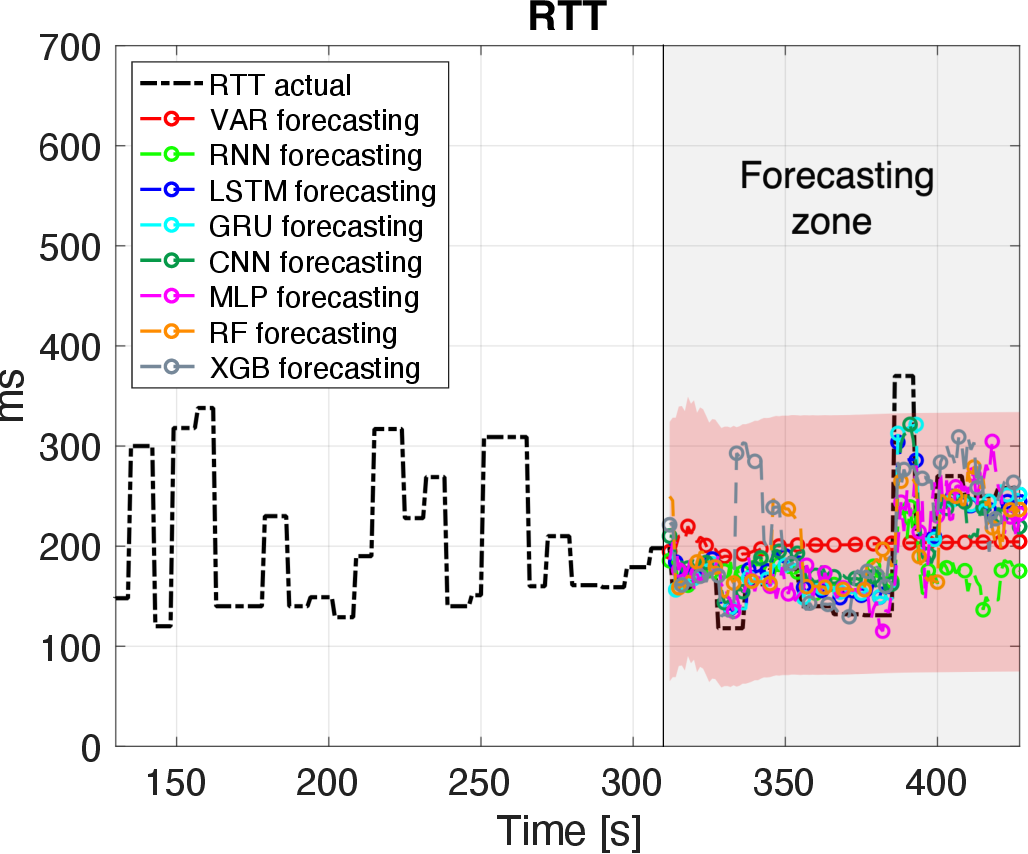}} \\
		\subfloat{\includegraphics[scale=0.33]{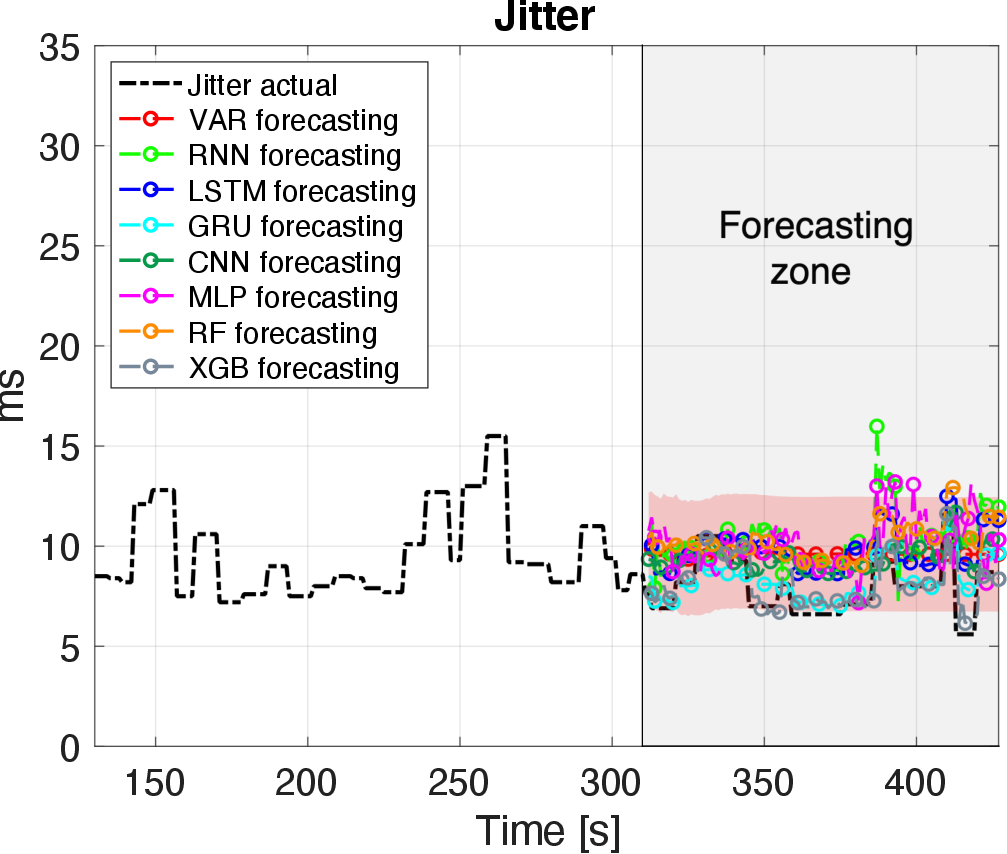}}  \hspace{2mm}
		\subfloat{\includegraphics[scale=0.33]{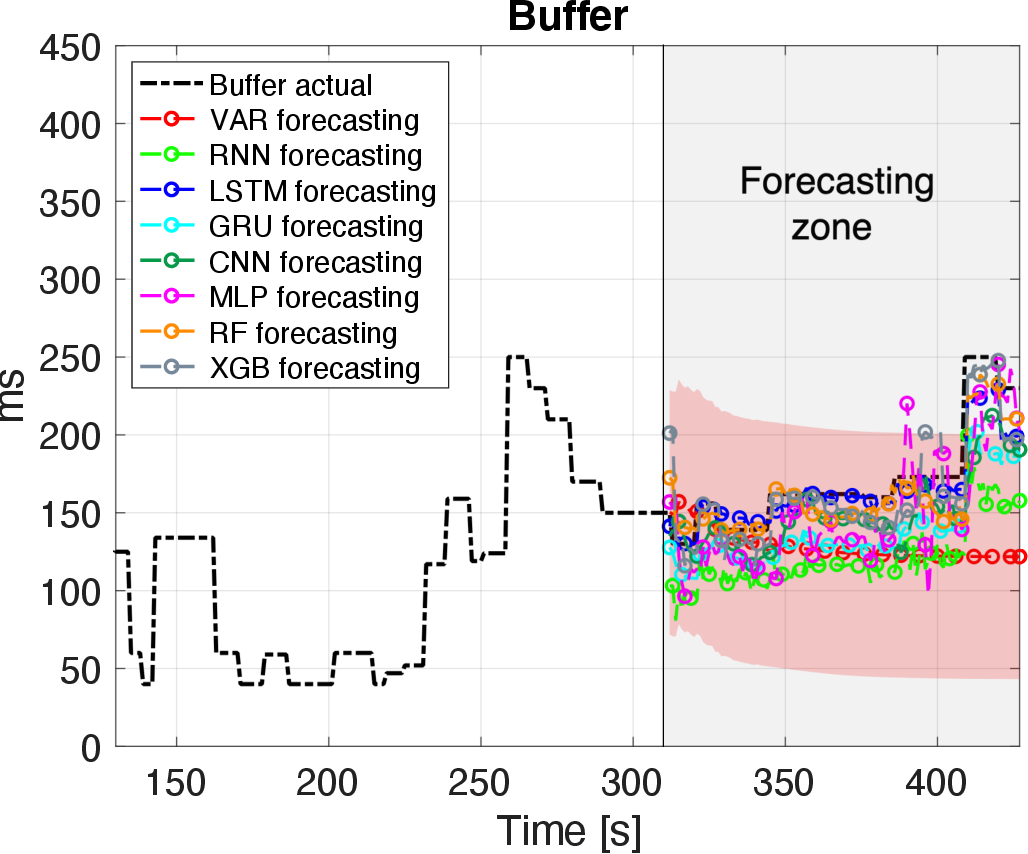}} \hspace{2mm}
		\subfloat{\includegraphics[scale=0.33]{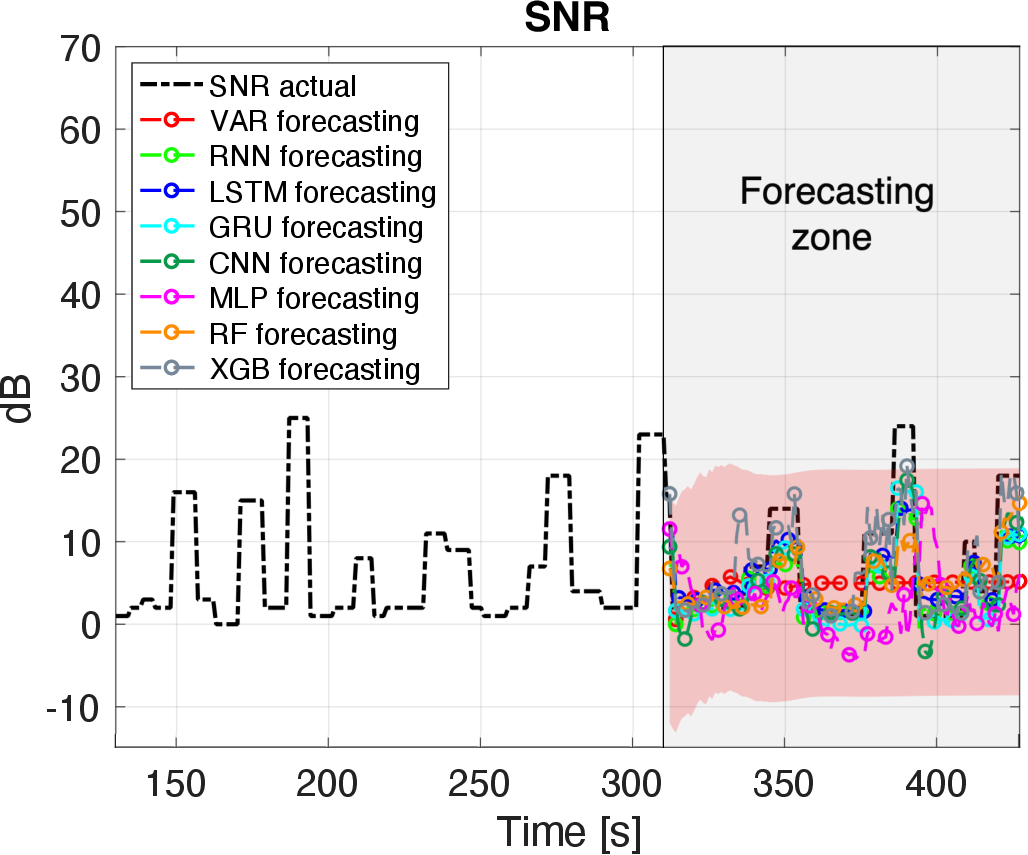}} 
	\end{tabular}
	\caption{G.729 codec: Multivariate time series forecasting for: MOS, Bandwidth, RTT, Jitter, Buffer, SNR, along with the 95\% forecasting intervals in shaded pale red. In the gray area on the right, the results of forecasting for each technique.}
	\label{fig:forecast_g729}
\end{figure*}

\begin{table*}[htp]
	\small
	\renewcommand{\arraystretch}{1.2}
	\centering
	\caption{Performance metrics (RMSE, MAE, MAPE) per codec (voice flows).}
	\resizebox{1\textwidth}{!}{%
    \begin{tabular}{ccccccc||ccccccc||ccccccc}
    	    \bottomrule
    \multicolumn{7}{c}{\textbf{Voice Track \#1 (G.722 codec)}} & \multicolumn{7}{c}{\textbf{Voice Track \#2 (G.729 codec)}} & \multicolumn{7}{c}{\textbf{Voice Track \#3 (MPEG-16 codec)}} \\
    \midrule
    \rowcolor[rgb]{ .906,  .902,  .902} \textbf{VAR(p*=12)} & \cellcolor[rgb]{ 1,  1,  1}\textbf{MOS} & \cellcolor[rgb]{ 1,  1,  1}\textbf{BW} & \cellcolor[rgb]{ 1,  1,  1}\textbf{RTT} & \cellcolor[rgb]{ 1,  1,  1}\textbf{Jitter} & \cellcolor[rgb]{ 1,  1,  1}\textbf{Buffer} & \cellcolor[rgb]{ 1,  1,  1}\textbf{SNR} & \textbf{VAR(p*=13)} & \cellcolor[rgb]{ 1,  1,  1}\textbf{MOS} & \cellcolor[rgb]{ 1,  1,  1}\textbf{BW} & \cellcolor[rgb]{ 1,  1,  1}\textbf{RTT} & \cellcolor[rgb]{ 1,  1,  1}\textbf{Jitter} & \cellcolor[rgb]{ 1,  1,  1}\textbf{Buffer} & \cellcolor[rgb]{ 1,  1,  1}\textbf{SNR} & \textbf{VAR(p*=11)} & \cellcolor[rgb]{ 1,  1,  1}\textbf{MOS} & \cellcolor[rgb]{ 1,  1,  1}\textbf{BW} & \cellcolor[rgb]{ 1,  1,  1}\textbf{RTT} & \cellcolor[rgb]{ 1,  1,  1}\textbf{Jitter} & \cellcolor[rgb]{ 1,  1,  1}\textbf{Buffer} & \cellcolor[rgb]{ 1,  1,  1}\textbf{SNR} \\
     RMSE & 0.017 & 14.01 & 10.41 & 32.52 & 48.74 & 11.41 &  RMSE & 0.011 & 3.57  & 66.2  & 21.99 & 58.1  & 7.27  &  RMSE & 0.006 & 3.97  & 349.9 & 34.07 & 63.12 & 9.17 \\
     MAE  & 0.0149 & 6.32  & 53.72 & 14.64 & 37.7  & 10.3  &  MAE  & 0.01  & 2.52  & 56.5  & 19.41 & 45.3  & 5.55  &  MAE  & 0.003 & 2.84  & 105.5 & 19.8  & 43.4  & 7.86 \\
    MAPE (\%) & 0.3   & \textcolor[rgb]{ 1,  0,  0}{\textit{8.0}} & 16.0  & 14.0  & 31.0  & 68.0  & MAPE (\%) & 0.2   & 17.0  & 32.0  & 27.0  & 24.0  & 165.0 & MAPE (\%) & 0.07  & 9.7   & 23.0  & 11.0  & 44.0  & 94.0 \\
    \rowcolor[rgb]{ .906,  .902,  .902} \textbf{RNN} & \cellcolor[rgb]{ 1,  1,  1}\textbf{MOS} & \cellcolor[rgb]{ 1,  1,  1}\textbf{BW} & \cellcolor[rgb]{ 1,  1,  1}\textbf{RTT} & \cellcolor[rgb]{ 1,  1,  1}\textbf{Jitter} & \cellcolor[rgb]{ 1,  1,  1}\textbf{Buffer} & \cellcolor[rgb]{ 1,  1,  1}\textbf{SNR} & \textbf{RNN} & \cellcolor[rgb]{ 1,  1,  1}\textbf{MOS} & \cellcolor[rgb]{ 1,  1,  1}\textbf{BW} & \cellcolor[rgb]{ 1,  1,  1}\textbf{RTT} & \cellcolor[rgb]{ 1,  1,  1}\textbf{Jitter} & \cellcolor[rgb]{ 1,  1,  1}\textbf{Buffer} & \cellcolor[rgb]{ 1,  1,  1}\textbf{SNR} & \textbf{RNN} & \cellcolor[rgb]{ 1,  1,  1}\textbf{MOS} & \cellcolor[rgb]{ 1,  1,  1}\textbf{BW} & \cellcolor[rgb]{ 1,  1,  1}\textbf{RTT} & \cellcolor[rgb]{ 1,  1,  1}\textbf{Jitter} & \cellcolor[rgb]{ 1,  1,  1}\textbf{Buffer} & \cellcolor[rgb]{ 1,  1,  1}\textbf{SNR} \\
     RMSE & 0.044 & 24.78 & 162.73 & 26.46 & 35.8  & 10.42 &  RMSE & 0.04  & 6.21  & 64.4  & 28.4  & 52.5  & 5.03  &  RMSE & 0.007 & 5.18  & 325.5 & 75.8  & 68.7  & 12.4 \\
     MAE  & 0.038 & 19.31 & 58.62 & 17.7  & 23.54 & 9.44  &  MAE  & 0.04  & 5.69  & 48.9  & 25.5  & 49.8  & 2.37  &  MAE  & 0.006 & 4.08  & 154.4 & 63.3  & 62.4  & 11.1 \\
    MAPE (\%) & 0.86  & 22.8  & 21.0  & 21.6  & 24.4  & 65.0  & MAPE (\%) & 1.0   & 26.0  & 23.0  & 35.0  & 29.0  & \textcolor[rgb]{ 1,  0,  0}{\textit{35.0}} & MAPE (\%) & 0.14  & 14.0  & 64.0  & 38.0  & 113.0 & 65.0 \\
    \rowcolor[rgb]{ .906,  .902,  .902} \textbf{LSTM} & \cellcolor[rgb]{ 1,  1,  1}\textbf{MOS} & \cellcolor[rgb]{ 1,  1,  1}\textbf{BW} & \cellcolor[rgb]{ 1,  1,  1}\textbf{RTT} & \cellcolor[rgb]{ 1,  1,  1}\textbf{Jitter} & \cellcolor[rgb]{ 1,  1,  1}\textbf{Buffer} & \cellcolor[rgb]{ 1,  1,  1}\textbf{SNR} & \textbf{LSTM} & \cellcolor[rgb]{ 1,  1,  1}\textbf{MOS} & \cellcolor[rgb]{ 1,  1,  1}\textbf{BW} & \cellcolor[rgb]{ 1,  1,  1}\textbf{RTT} & \cellcolor[rgb]{ 1,  1,  1}\textbf{Jitter} & \cellcolor[rgb]{ 1,  1,  1}\textbf{Buffer} & \cellcolor[rgb]{ 1,  1,  1}\textbf{SNR} & \textbf{LSTM} & \cellcolor[rgb]{ 1,  1,  1}\textbf{MOS} & \cellcolor[rgb]{ 1,  1,  1}\textbf{BW} & \cellcolor[rgb]{ 1,  1,  1}\textbf{RTT} & \cellcolor[rgb]{ 1,  1,  1}\textbf{Jitter} & \cellcolor[rgb]{ 1,  1,  1}\textbf{Buffer} & \cellcolor[rgb]{ 1,  1,  1}\textbf{SNR} \\
     RMSE & 0.016 & 14.05 & 142.21 & 16.29 & 33.73 & 5.73  &  RMSE & 0.065 & 3.97  & 33.69 & 21.8  & 14.8  & 4.58  &  RMSE & 0.006 & 4.23  & 229.15 & 32.79 & 41.76 & 6.31 \\
     MAE  & 0.007 & 6.78  & 53.25 & 7.79  & 20.13 & 4.92  &  MAE  & 0.063 & 3.04  & 22.19 & 19.2  & 9.43  & 3.21  &  MAE  & 0.003 & 2.75  & 67.1  & 20.83 & 18.83 & 3.55 \\
    MAPE (\%) & \textcolor[rgb]{ 1,  0,  0}{\textit{0.16}} & 9.3   & 22.9  & 8.8   & 15.7  & 39.0  & MAPE (\%) & 1.4   & 18.0  & 11.0  & 26.0  & \textcolor[rgb]{ 1,  0,  0}{\textit{4.9}} & 90.0  & MAPE (\%) & 0.07  & \textcolor[rgb]{ 1,  0,  0}{\textit{9.0}} & 16.4  & 12.0  & 14.0  & 88.0 \\
    \rowcolor[rgb]{ .906,  .902,  .902} \textbf{GRU} & \cellcolor[rgb]{ 1,  1,  1}\textbf{MOS} & \cellcolor[rgb]{ 1,  1,  1}\textbf{BW} & \cellcolor[rgb]{ 1,  1,  1}\textbf{RTT} & \cellcolor[rgb]{ 1,  1,  1}\textbf{Jitter} & \cellcolor[rgb]{ 1,  1,  1}\textbf{Buffer} & \cellcolor[rgb]{ 1,  1,  1}\textbf{SNR} & \textbf{GRU} & \cellcolor[rgb]{ 1,  1,  1}\textbf{MOS} & \cellcolor[rgb]{ 1,  1,  1}\textbf{BW} & \cellcolor[rgb]{ 1,  1,  1}\textbf{RTT} & \cellcolor[rgb]{ 1,  1,  1}\textbf{Jitter} & \cellcolor[rgb]{ 1,  1,  1}\textbf{Buffer} & \cellcolor[rgb]{ 1,  1,  1}\textbf{SNR} & \textbf{GRU} & \cellcolor[rgb]{ 1,  1,  1}\textbf{MOS} & \cellcolor[rgb]{ 1,  1,  1}\textbf{BW} & \cellcolor[rgb]{ 1,  1,  1}\textbf{RTT} & \cellcolor[rgb]{ 1,  1,  1}\textbf{Jitter} & \cellcolor[rgb]{ 1,  1,  1}\textbf{Buffer} & \cellcolor[rgb]{ 1,  1,  1}\textbf{SNR} \\
     RMSE & 0.016 & 14.04 & 148.67 & 16.94 & 24.05 & 4.04  &  RMSE & 0.04  & 5.24  & 31.36 & 11.5  & 35.8  & 4.71  &  RMSE & 0.004 & 4.39  & 235.75 & 27.1  & 34.52 & 5.79 \\
     MAE  & 0.011 & 6.82  & 52.37 & 7.34  & 13.25 & 3.14  &  MAE  & 0.038 & 4.62  & 18.7  & 8.25  & 33.17 & 3.22  &  MAE  & 0.003 & 3.1   & 74.5  & 15.08 & 18.46 & 3.97 \\
    MAPE (\%) & 0.26  & 9.4   & 20.5  & \textcolor[rgb]{ 1,  0,  0}{\textit{7.3}} & 11.0  & \textcolor[rgb]{ 1,  0,  0}{\textit{31.0}} & MAPE (\%) & 0.9   & 22.0  & \textcolor[rgb]{ 1,  0,  0}{\textit{10.0}} & 11.0  & 19.0  & 69.0  & MAPE (\%) & 0.06  & 9.6   & 20.0  & 8.0   & 15.0  & 79.0 \\
    \rowcolor[rgb]{ .906,  .902,  .902} \textbf{CNN} & \cellcolor[rgb]{ 1,  1,  1}\textbf{MOS} & \cellcolor[rgb]{ 1,  1,  1}\textbf{BW} & \cellcolor[rgb]{ 1,  1,  1}\textbf{RTT} & \cellcolor[rgb]{ 1,  1,  1}\textbf{Jitter} & \cellcolor[rgb]{ 1,  1,  1}\textbf{Buffer} & \cellcolor[rgb]{ 1,  1,  1}\textbf{SNR} & \textbf{CNN} & \cellcolor[rgb]{ 1,  1,  1}\textbf{MOS} & \cellcolor[rgb]{ 1,  1,  1}\textbf{BW} & \cellcolor[rgb]{ 1,  1,  1}\textbf{RTT} & \cellcolor[rgb]{ 1,  1,  1}\textbf{Jitter} & \cellcolor[rgb]{ 1,  1,  1}\textbf{Buffer} & \cellcolor[rgb]{ 1,  1,  1}\textbf{SNR} & \textbf{CNN} & \cellcolor[rgb]{ 1,  1,  1}\textbf{MOS} & \cellcolor[rgb]{ 1,  1,  1}\textbf{BW} & \cellcolor[rgb]{ 1,  1,  1}\textbf{RTT} & \cellcolor[rgb]{ 1,  1,  1}\textbf{Jitter} & \cellcolor[rgb]{ 1,  1,  1}\textbf{Buffer} & \cellcolor[rgb]{ 1,  1,  1}\textbf{SNR} \\
     RMSE & 0.03  & 14.36 & 145.84 & 22.86 & 27.06 & 5.88  &  RMSE & 0.024 & 3.25  & 37.65 & 18.96 & 27.6  & 4.63  &  RMSE & 0.0045 & 4.35  & 198.99 & 25.54 & 21.4  & 5.82 \\
     MAE  & 0.017 & 7.26  & 48.78 & 11.36 & 16.97 & 5.16  &  MAE  & 0.018 & 1.8   & 30.31 & 16.3  & 22.36 & 3.21  &  MAE  & 0.0026 & 2.86  & 66.4  & 14.45 & 13.45 & 3.79 \\
    MAPE (\%) & 0.4   & 9.7   & 19.0  & 11.8  & 14.3  & 42.0  & MAPE (\%) & 0.4   & \textcolor[rgb]{ 1,  0,  0}{\textit{14.0}} & 17.0  & 23.0  & 12.0  & 80.0  & MAPE (\%) & 0.061 & 9.4   & 23.0  & 6.7   & 15.0  & 72.0 \\
    \rowcolor[rgb]{ .906,  .902,  .902} \textbf{MLP} & \cellcolor[rgb]{ 1,  1,  1}\textbf{MOS} & \cellcolor[rgb]{ 1,  1,  1}\textbf{BW} & \cellcolor[rgb]{ 1,  1,  1}\textbf{RTT} & \cellcolor[rgb]{ 1,  1,  1}\textbf{Jitter} & \cellcolor[rgb]{ 1,  1,  1}\textbf{Buffer} & \cellcolor[rgb]{ 1,  1,  1}\textbf{SNR} & \textbf{MLP} & \cellcolor[rgb]{ 1,  1,  1}\textbf{MOS} & \cellcolor[rgb]{ 1,  1,  1}\textbf{BW} & \cellcolor[rgb]{ 1,  1,  1}\textbf{RTT} & \cellcolor[rgb]{ 1,  1,  1}\textbf{Jitter} & \cellcolor[rgb]{ 1,  1,  1}\textbf{Buffer} & \cellcolor[rgb]{ 1,  1,  1}\textbf{SNR} & \textbf{MLP} & \cellcolor[rgb]{ 1,  1,  1}\textbf{MOS} & \cellcolor[rgb]{ 1,  1,  1}\textbf{BW} & \cellcolor[rgb]{ 1,  1,  1}\textbf{RTT} & \cellcolor[rgb]{ 1,  1,  1}\textbf{Jitter} & \cellcolor[rgb]{ 1,  1,  1}\textbf{Buffer} & \cellcolor[rgb]{ 1,  1,  1}\textbf{SNR} \\
     RMSE & 0.03  & 50.32 & 181.8 & 33.06 & 68.10 & 8.63  &  RMSE & 0.05  & 3.59  & 57.5  & 22.79 & 30.62 & 4.61  &  RMSE & 0.004 & 4.86  & 231.6 & 42.13 & 39.4  & 8.25 \\
     MAE  & 0.02  & 28.81 & 81.09 & 18.14 & 41.97 & 6.57  &  MAE  & 0.05  & 2.19  & 49.8  & 20.23 & 25.7  & 3.44  &  MAE  & 0.0025 & 3.33  & 74.3  & 21.3  & 23.4  & 2.7 \\
    MAPE (\%) & 0.4   & 35.0  & 40.0  & 19.0  & 34.0  & 54.0  & MAPE (\%) & 1.1   & 17.0  & 30.0  & 28.0  & 15.0  & 118.0 & MAPE (\%) & 0.06  & 11.0  & 21.0  & 12.0  & 22.0  & \textcolor[rgb]{ 1,  0,  0}{\textit{49.0}} \\
    \rowcolor[rgb]{ .906,  .902,  .902} \textbf{RF} & \cellcolor[rgb]{ 1,  1,  1}\textbf{MOS} & \cellcolor[rgb]{ 1,  1,  1}\textbf{BW} & \cellcolor[rgb]{ 1,  1,  1}\textbf{RTT} & \cellcolor[rgb]{ 1,  1,  1}\textbf{Jitter} & \cellcolor[rgb]{ 1,  1,  1}\textbf{Buffer} & \cellcolor[rgb]{ 1,  1,  1}\textbf{SNR} & \textbf{RF} & \cellcolor[rgb]{ 1,  1,  1}\textbf{MOS} & \cellcolor[rgb]{ 1,  1,  1}\textbf{BW} & \cellcolor[rgb]{ 1,  1,  1}\textbf{RTT} & \cellcolor[rgb]{ 1,  1,  1}\textbf{Jitter} & \cellcolor[rgb]{ 1,  1,  1}\textbf{Buffer} & \cellcolor[rgb]{ 1,  1,  1}\textbf{SNR} & \textbf{RF} & \cellcolor[rgb]{ 1,  1,  1}\textbf{MOS} & \cellcolor[rgb]{ 1,  1,  1}\textbf{BW} & \cellcolor[rgb]{ 1,  1,  1}\textbf{RTT} & \cellcolor[rgb]{ 1,  1,  1}\textbf{Jitter} & \cellcolor[rgb]{ 1,  1,  1}\textbf{Buffer} & \cellcolor[rgb]{ 1,  1,  1}\textbf{SNR} \\
     RMSE & 0.025 & 15.15 & 167.39 & 29.64 & 33.39 & 8.03  &  RMSE & 0.05  & 5.14  & 45.65 & 27.56 & 17.25 & 5.41  &  RMSE & 0.006 & 4.38  & 332.7 & 30.5  & 23.3  & 5.93 \\
     MAE  & 0.023 & 7.6   & 53.08 & 18.5  & 23.65 & 6.27  &  MAE  & 0.043 & 3.87  & 34.97 & 25.16 & 12.41 & 3.89  &  MAE  & 0.002 & 2.8   & 93.1  & 18.6  & 15.2  & 4.15 \\
    MAPE (\%) & 0.52  & 10.0  & 17.0  & 22.0  & 20.0  & 50.0  & MAPE (\%) & 1.0   & 21.0  & 19.0  & 34.0  & 7.0   & 112.0 & MAPE (\%) & 0.05  & 9.0   & 17.0  & 10.0  & 17.0  & 54.0 \\
    \rowcolor[rgb]{ .906,  .902,  .902} \textbf{XGB} & \cellcolor[rgb]{ 1,  1,  1}\textbf{MOS} & \cellcolor[rgb]{ 1,  1,  1}\textbf{BW} & \cellcolor[rgb]{ 1,  1,  1}\textbf{RTT} & \cellcolor[rgb]{ 1,  1,  1}\textbf{Jitter} & \cellcolor[rgb]{ 1,  1,  1}\textbf{Buffer} & \cellcolor[rgb]{ 1,  1,  1}\textbf{SNR} & \textbf{XGB} & \cellcolor[rgb]{ 1,  1,  1}\textbf{MOS} & \cellcolor[rgb]{ 1,  1,  1}\textbf{BW} & \cellcolor[rgb]{ 1,  1,  1}\textbf{RTT} & \cellcolor[rgb]{ 1,  1,  1}\textbf{Jitter} & \cellcolor[rgb]{ 1,  1,  1}\textbf{Buffer} & \cellcolor[rgb]{ 1,  1,  1}\textbf{SNR} & \textbf{XGB} & \cellcolor[rgb]{ 1,  1,  1}\textbf{MOS} & \cellcolor[rgb]{ 1,  1,  1}\textbf{BW} & \cellcolor[rgb]{ 1,  1,  1}\textbf{RTT} & \cellcolor[rgb]{ 1,  1,  1}\textbf{Jitter} & \cellcolor[rgb]{ 1,  1,  1}\textbf{Buffer} & \cellcolor[rgb]{ 1,  1,  1}\textbf{SNR} \\
    RMSE  & 0.017 & 14.41 & 160.57 & 29.28 & 20.78 & 6.28  & RMSE  & 0.010 & 4.57  & 56.9  & 9.45  & 19.75 & 4.18  & RMSE  & 0.007 & 4.51  & 270.72 & 22.6  & 17.75 & 7.32 \\
    MAE   & 0.012 & 6.55  & 47.1  & 13.77 & 11.21 & 4.46  & MAE   & 0.004 & 3.2   & 37.5  & 5.75  & 14.18 & 2.57  & MAE   & 0.0019 & 3.01  & 60.1  & 1.24  & 8.91  & 4.62 \\
    MAPE (\%) & 0.27  & 9.5   & \textcolor[rgb]{ 1,  0,  0}{\textit{14.5}} & 13.0  & \textcolor[rgb]{ 1,  0,  0}{\textit{10.0}} & 37.0  & MAPE (\%) & \textcolor[rgb]{ 1,  0,  0}{\textit{0.1}} & 17.0  & 21.0  & \textcolor[rgb]{ 1,  0,  0}{7.0} & 8.0   & 82.0  & MAPE (\%) & \textcolor[rgb]{ 1,  0,  0}{\textit{0.04}} & 9.8   & \textcolor[rgb]{ 1,  0,  0}{\textit{11.0}} & \textcolor[rgb]{ 1,  0,  0}{5.0} & \textcolor[rgb]{ 1,  0,  0}{9.2} & 58.0 \\    \bottomrule
    \bottomrule
    \multicolumn{7}{c}{\textbf{Voice Track \#4 (OPUS codec)}} & \multicolumn{7}{c}{\textbf{Voice Track \#5 (GSM codec)}} & \multicolumn{7}{c}{\textbf{Voice Track \#6 (SPX8000 codec)}} \\
    \midrule
    \rowcolor[rgb]{ .906,  .902,  .902} \textbf{VAR(p*=11)} & \cellcolor[rgb]{ 1,  1,  1}\textbf{MOS} & \cellcolor[rgb]{ 1,  1,  1}\textbf{BW} & \cellcolor[rgb]{ 1,  1,  1}\textbf{RTT} & \cellcolor[rgb]{ 1,  1,  1}\textbf{Jitter} & \cellcolor[rgb]{ 1,  1,  1}\textbf{Buffer} & \cellcolor[rgb]{ 1,  1,  1}\textbf{SNR} & \textbf{VAR(p*=5)} & \cellcolor[rgb]{ 1,  1,  1}\textbf{MOS} & \cellcolor[rgb]{ 1,  1,  1}\textbf{BW} & \cellcolor[rgb]{ 1,  1,  1}\textbf{RTT} & \cellcolor[rgb]{ 1,  1,  1}\textbf{Jitter} & \cellcolor[rgb]{ 1,  1,  1}\textbf{Buffer} & \cellcolor[rgb]{ 1,  1,  1}\textbf{SNR} & \textbf{VAR(p*=8)} & \cellcolor[rgb]{ 1,  1,  1}\textbf{MOS} & \cellcolor[rgb]{ 1,  1,  1}\textbf{BW} & \cellcolor[rgb]{ 1,  1,  1}\textbf{RTT} & \cellcolor[rgb]{ 1,  1,  1}\textbf{Jitter} & \cellcolor[rgb]{ 1,  1,  1}\textbf{Buffer} & \cellcolor[rgb]{ 1,  1,  1}\textbf{SNR} \\
     RMSE & 0.02  & 3.2   & 22.2  & 67.9  & 78.1  & 6.67  &  RMSE & 0.06  & 3.7   & 37.5  & 7.6   & 60.0  & 10.6  &  RMSE & 0.02  & 4.9   & 108.4 & 30.2  & 123.3 & 8.7 \\
     MAE  & 0.02  & 1.58  & 18.2  & 60.1  & 77.3  & 5.01  &  MAE  & 0.06  & 1.5   & 21.5  & 6.3   & 42.8  & 3.3   &  MAE  & 0.014 & 3.9   & 56.8  & 18.5  & 109.2 & 6.4 \\
    MAPE (\%) & 0.4   & 3.1   & 15.0  & 13.0  & 177.0 & 48.0  & MAPE (\%) & 1.3   & 25.0  & 13.0  & 8.0   & 38.0  & 170.0 & MAPE (\%) & 0.3   & 13.0  & 23.0  & 17.0  & 58.0  & 26.0 \\
    \rowcolor[rgb]{ .906,  .902,  .902} \textbf{RNN} & \cellcolor[rgb]{ 1,  1,  1}\textbf{MOS} & \cellcolor[rgb]{ 1,  1,  1}\textbf{BW} & \cellcolor[rgb]{ 1,  1,  1}\textbf{RTT} & \cellcolor[rgb]{ 1,  1,  1}\textbf{Jitter} & \cellcolor[rgb]{ 1,  1,  1}\textbf{Buffer} & \cellcolor[rgb]{ 1,  1,  1}\textbf{SNR} & \textbf{RNN} & \cellcolor[rgb]{ 1,  1,  1}\textbf{MOS} & \cellcolor[rgb]{ 1,  1,  1}\textbf{BW} & \cellcolor[rgb]{ 1,  1,  1}\textbf{RTT} & \cellcolor[rgb]{ 1,  1,  1}\textbf{Jitter} & \cellcolor[rgb]{ 1,  1,  1}\textbf{Buffer} & \cellcolor[rgb]{ 1,  1,  1}\textbf{SNR} & \textbf{RNN} & \cellcolor[rgb]{ 1,  1,  1}\textbf{MOS} & \cellcolor[rgb]{ 1,  1,  1}\textbf{BW} & \cellcolor[rgb]{ 1,  1,  1}\textbf{RTT} & \cellcolor[rgb]{ 1,  1,  1}\textbf{Jitter} & \cellcolor[rgb]{ 1,  1,  1}\textbf{Buffer} & \cellcolor[rgb]{ 1,  1,  1}\textbf{SNR} \\
     RMSE & 0.04  & 3.25  & 37.8  & 201.1 & 40.35 & 5.9   &  RMSE & 0.06  & 12.8  & 46.8  & 22.0  & 35.7  & 9.26  &  RMSE & 0.06  & 5.14  & 110.0 & 47.6  & 110.7 & 8.01 \\
     MAE  & 0.04  & 1.62  & 35.8  & 198.1 & 38.9  & 4.9   &  MAE  & 0.05  & 11.8  & 43.2  & 21.6  & 26.5  & 8.3   &  MAE  & 0.06  & 4.2   & 50.2  & 40.1  & 96.4  & 6.34 \\
    MAPE (\%) & 0.8   & 3.2   & 28.0  & 42.0  & 90.0  & 61.0  & MAPE (\%) & 1.0   & 40.0  & 30.0  & 29.0  & 30.0  & 260.0 & MAPE (\%) & 1.4   & 14.0  & 17.0  & 40.0  & 50.0  & 28.0 \\
    \rowcolor[rgb]{ .906,  .902,  .902} \textbf{LSTM} & \cellcolor[rgb]{ 1,  1,  1}\textbf{MOS} & \cellcolor[rgb]{ 1,  1,  1}\textbf{BW} & \cellcolor[rgb]{ 1,  1,  1}\textbf{RTT} & \cellcolor[rgb]{ 1,  1,  1}\textbf{Jitter} & \cellcolor[rgb]{ 1,  1,  1}\textbf{Buffer} & \cellcolor[rgb]{ 1,  1,  1}\textbf{SNR} & \textbf{LSTM} & \cellcolor[rgb]{ 1,  1,  1}\textbf{MOS} & \cellcolor[rgb]{ 1,  1,  1}\textbf{BW} & \cellcolor[rgb]{ 1,  1,  1}\textbf{RTT} & \cellcolor[rgb]{ 1,  1,  1}\textbf{Jitter} & \cellcolor[rgb]{ 1,  1,  1}\textbf{Buffer} & \cellcolor[rgb]{ 1,  1,  1}\textbf{SNR} & \textbf{LSTM} & \cellcolor[rgb]{ 1,  1,  1}\textbf{MOS} & \cellcolor[rgb]{ 1,  1,  1}\textbf{BW} & \cellcolor[rgb]{ 1,  1,  1}\textbf{RTT} & \cellcolor[rgb]{ 1,  1,  1}\textbf{Jitter} & \cellcolor[rgb]{ 1,  1,  1}\textbf{Buffer} & \cellcolor[rgb]{ 1,  1,  1}\textbf{SNR} \\
     RMSE & 0.016 & 3.25  & 8.05  & 27.97 & 16.73 & 3.96  &  RMSE & 0.06  & 3.8   & 35.1  & 4.4   & 24.4  & 5.4   &  RMSE & 0.016 & 4.9   & 80.8  & 16.7  & 63.7  & 6.5 \\
     MAE  & 0.016 & 1.53  & 6.4   & 22.8  & 15.63 & 2.78  &  MAE  & 0.06  & 1.9   & 20.5  & 3.4   & 14.4  & 3.9   &  MAE  & 0.013 & 3.8   & 30.8  & 10.5  & 45.8  & 4.6 \\
    MAPE (\%) & 0.37  & 3.2   & 5.0   & 5.0   & 36.0  & \textcolor[rgb]{ 1,  0,  0}{\textit{32.0}} & MAPE (\%) & 1.3   & 25.0  & 12.0  & 4.0   & 15.0  & 80.0  & MAPE (\%) & 0.3   & \textcolor[rgb]{ 1,  0,  0}{\textit{12.0}} & \textcolor[rgb]{ 1,  0,  0}{\textit{14.0}} & 10.0  & 20.0  & 20.0 \\
    \rowcolor[rgb]{ .906,  .902,  .902} \textbf{GRU} & \cellcolor[rgb]{ 1,  1,  1}\textbf{MOS} & \cellcolor[rgb]{ 1,  1,  1}\textbf{BW} & \cellcolor[rgb]{ 1,  1,  1}\textbf{RTT} & \cellcolor[rgb]{ 1,  1,  1}\textbf{Jitter} & \cellcolor[rgb]{ 1,  1,  1}\textbf{Buffer} & \cellcolor[rgb]{ 1,  1,  1}\textbf{SNR} & \textbf{GRU} & \cellcolor[rgb]{ 1,  1,  1}\textbf{MOS} & \cellcolor[rgb]{ 1,  1,  1}\textbf{BW} & \cellcolor[rgb]{ 1,  1,  1}\textbf{RTT} & \cellcolor[rgb]{ 1,  1,  1}\textbf{Jitter} & \cellcolor[rgb]{ 1,  1,  1}\textbf{Buffer} & \cellcolor[rgb]{ 1,  1,  1}\textbf{SNR} & \textbf{GRU} & \cellcolor[rgb]{ 1,  1,  1}\textbf{MOS} & \cellcolor[rgb]{ 1,  1,  1}\textbf{BW} & \cellcolor[rgb]{ 1,  1,  1}\textbf{RTT} & \cellcolor[rgb]{ 1,  1,  1}\textbf{Jitter} & \cellcolor[rgb]{ 1,  1,  1}\textbf{Buffer} & \cellcolor[rgb]{ 1,  1,  1}\textbf{SNR} \\
     RMSE & 0.005 & 3.19  & 19.36 & 17.7  & 36.5  & 5.57  &  RMSE & 0.04  & 3.8   & 33.5  & 4.1   & 20.4  & 4.9   &  RMSE & 0.01  & 5.01  & 68.6  & 17.4  & 49.7  & 7.1 \\
     MAE  & 0.005 & 1.3   & 18.6  & 12.6  & 35.6  & 4.8   &  MAE  & 0.03  & 2.03  & 20.7  & 3.1   & 1.4   & 2.8   &  MAE  & 0.008 & 4.04  & 41.3  & 10.4  & 40.5  & 5.2 \\
    MAPE (\%) & 0.1   & \textcolor[rgb]{ 1,  0,  0}{\textit{3.0}} & 15.0  & \textcolor[rgb]{ 1,  0,  0}{\textit{2.8}} & 82.0  & 58.0  & MAPE (\%) & 0.8   & 25.0  & 13.0  & 4.0   & \textcolor[rgb]{ 1,  0,  0}{\textit{10.0}} & \textcolor[rgb]{ 1,  0,  0}{\textit{48.0}} & MAPE (\%) & 0.2   & 13.0  & 20.0  & 9.6   & 20.0  & 26.0 \\
    \rowcolor[rgb]{ .906,  .902,  .902} \textbf{CNN} & \cellcolor[rgb]{ 1,  1,  1}\textbf{MOS} & \cellcolor[rgb]{ 1,  1,  1}\textbf{BW} & \cellcolor[rgb]{ 1,  1,  1}\textbf{RTT} & \cellcolor[rgb]{ 1,  1,  1}\textbf{Jitter} & \cellcolor[rgb]{ 1,  1,  1}\textbf{Buffer} & \cellcolor[rgb]{ 1,  1,  1}\textbf{SNR} & \textbf{CNN} & \cellcolor[rgb]{ 1,  1,  1}\textbf{MOS} & \cellcolor[rgb]{ 1,  1,  1}\textbf{BW} & \cellcolor[rgb]{ 1,  1,  1}\textbf{RTT} & \cellcolor[rgb]{ 1,  1,  1}\textbf{Jitter} & \cellcolor[rgb]{ 1,  1,  1}\textbf{Buffer} & \cellcolor[rgb]{ 1,  1,  1}\textbf{SNR} & \textbf{CNN} & \cellcolor[rgb]{ 1,  1,  1}\textbf{MOS} & \cellcolor[rgb]{ 1,  1,  1}\textbf{BW} & \cellcolor[rgb]{ 1,  1,  1}\textbf{RTT} & \cellcolor[rgb]{ 1,  1,  1}\textbf{Jitter} & \cellcolor[rgb]{ 1,  1,  1}\textbf{Buffer} & \cellcolor[rgb]{ 1,  1,  1}\textbf{SNR} \\
     RMSE & 0.008 & 3.44  & 10.1  & 23.9  & 13.1  & 4.38  &  RMSE & 0.021 & 4.6   & 22.4  & 3.48  & 24.3  & 4.8   &  RMSE & 0.02  & 5.26  & 55.01 & 13.3  & 26.3  & 6.78 \\
     MAE  & 0.008 & 1.84  & 8.25  & 19.6  & 11.5  & 3.32  &  MAE  & 0.018 & 3.0   & 15.1  & 2.72  & 11.5  & 3.05  &  MAE  & 0.02  & 4.14  & 33.2  & 8.6   & 18.4  & 5.3 \\
    MAPE (\%) & 0.2   & 4.0   & 6.0   & 4.0   & 26.0  & 42.0  & MAPE (\%) & 0.4   & 30.0  & 8.0   & 3.7   & 11.0  & 70.0  & MAPE (\%) & 0.2   & 13.0  & 17.0  & \textcolor[rgb]{ 1,  0,  0}{\textit{9.0}} & \textcolor[rgb]{ 1,  0,  0}{\textit{9.7}} & 26.0 \\
    \rowcolor[rgb]{ .906,  .902,  .902} \textbf{MLP} & \cellcolor[rgb]{ 1,  1,  1}\textbf{MOS} & \cellcolor[rgb]{ 1,  1,  1}\textbf{BW} & \cellcolor[rgb]{ 1,  1,  1}\textbf{RTT} & \cellcolor[rgb]{ 1,  1,  1}\textbf{Jitter} & \cellcolor[rgb]{ 1,  1,  1}\textbf{Buffer} & \cellcolor[rgb]{ 1,  1,  1}\textbf{SNR} & \textbf{MLP} & \cellcolor[rgb]{ 1,  1,  1}\textbf{MOS} & \cellcolor[rgb]{ 1,  1,  1}\textbf{BW} & \cellcolor[rgb]{ 1,  1,  1}\textbf{RTT} & \cellcolor[rgb]{ 1,  1,  1}\textbf{Jitter} & \cellcolor[rgb]{ 1,  1,  1}\textbf{Buffer} & \cellcolor[rgb]{ 1,  1,  1}\textbf{SNR} & \textbf{MLP} & \cellcolor[rgb]{ 1,  1,  1}\textbf{MOS} & \cellcolor[rgb]{ 1,  1,  1}\textbf{BW} & \cellcolor[rgb]{ 1,  1,  1}\textbf{RTT} & \cellcolor[rgb]{ 1,  1,  1}\textbf{Jitter} & \cellcolor[rgb]{ 1,  1,  1}\textbf{Buffer} & \cellcolor[rgb]{ 1,  1,  1}\textbf{SNR} \\
     RMSE & 0.01  & 4.6   & 10.4  & 34.3  & 23.05 & 5.03  &  RMSE & 0.06  & 4.1   & 32.3  & 3.4   & 21.6  & 4.9   &  RMSE & 0.02  & 5.1   & 51.6  & 15.3  & 35.4  & 7.2 \\
     MAE  & 0.01  & 3.1   & 8.32  & 29.3  & 21.37 & 4.09  &  MAE  & 0.04  & 2.3   & 19.2  & 2.5   & 10.6  & 3.1   &  MAE  & 0.014 & 4.14  & 31.26 & 9.9   & 27.1  & 5.7 \\
    MAPE (\%) & 0.2   & 6.0   & 6.0   & 6.0   & 48.0  & 50.0  & MAPE (\%) & 0.9   & 25.0  & 12.0   & 3.65   & 11.0  & 70.0  & MAPE (\%) & 0.3   & 13.0  & 16.0  & 11.0  & 15.0  & 26.0 \\
    \rowcolor[rgb]{ .906,  .902,  .902} \textbf{RF} & \cellcolor[rgb]{ 1,  1,  1}\textbf{MOS} & \cellcolor[rgb]{ 1,  1,  1}\textbf{BW} & \cellcolor[rgb]{ 1,  1,  1}\textbf{RTT} & \cellcolor[rgb]{ 1,  1,  1}\textbf{Jitter} & \cellcolor[rgb]{ 1,  1,  1}\textbf{Buffer} & \cellcolor[rgb]{ 1,  1,  1}\textbf{SNR} & \textbf{RF} & \cellcolor[rgb]{ 1,  1,  1}\textbf{MOS} & \cellcolor[rgb]{ 1,  1,  1}\textbf{BW} & \cellcolor[rgb]{ 1,  1,  1}\textbf{RTT} & \cellcolor[rgb]{ 1,  1,  1}\textbf{Jitter} & \cellcolor[rgb]{ 1,  1,  1}\textbf{Buffer} & \cellcolor[rgb]{ 1,  1,  1}\textbf{SNR} & \textbf{RF} & \cellcolor[rgb]{ 1,  1,  1}\textbf{MOS} & \cellcolor[rgb]{ 1,  1,  1}\textbf{BW} & \cellcolor[rgb]{ 1,  1,  1}\textbf{RTT} & \cellcolor[rgb]{ 1,  1,  1}\textbf{Jitter} & \cellcolor[rgb]{ 1,  1,  1}\textbf{Buffer} & \cellcolor[rgb]{ 1,  1,  1}\textbf{SNR} \\
     RMSE & 0.014 & 3.05  & 8.32  & 69.3  & 19.07 & 4.5   &  RMSE & 0.06  & 3.7   & 36.3  & 6.0   & 42.2  & 5.1   &  RMSE & 0.02  & 5.2   & 95.5  & 21.9  & 80.7  & 6.8 \\
     MAE  & 0.013 & 1.5   & 6.1   & 54.2  & 16.4  & 2.97  &  MAE  & 0.05  & 1.4   & 21.4  & 4.6   & 27.9  & 3.3   &  MAE  & 0.02  & 4.01  & 54.8  & 15.1  & 66.1  & 5.6 \\
    MAPE (\%) & 0.3   & 3.0   & 4.8   & 11.0  & 38.0  & 35.0  & MAPE (\%) & 1.0   & \textcolor[rgb]{ 1,  0,  0}{\textit{20.0}} & 13.0  & 6.0   & 24.0  & 65.0  & MAPE (\%) & 0.4   & 13.0  & 24.0  & 16.0  & 33.0  & 26.0 \\
    \rowcolor[rgb]{ .906,  .902,  .902} \textbf{XGB} & \cellcolor[rgb]{ 1,  1,  1}\textbf{MOS} & \cellcolor[rgb]{ 1,  1,  1}\textbf{BW} & \cellcolor[rgb]{ 1,  1,  1}\textbf{RTT} & \cellcolor[rgb]{ 1,  1,  1}\textbf{Jitter} & \cellcolor[rgb]{ 1,  1,  1}\textbf{Buffer} & \cellcolor[rgb]{ 1,  1,  1}\textbf{SNR} & \textbf{XGB} & \cellcolor[rgb]{ 1,  1,  1}\textbf{MOS} & \cellcolor[rgb]{ 1,  1,  1}\textbf{BW} & \cellcolor[rgb]{ 1,  1,  1}\textbf{RTT} & \cellcolor[rgb]{ 1,  1,  1}\textbf{Jitter} & \cellcolor[rgb]{ 1,  1,  1}\textbf{Buffer} & \cellcolor[rgb]{ 1,  1,  1}\textbf{SNR} & \textbf{XGB} & \cellcolor[rgb]{ 1,  1,  1}\textbf{MOS} & \cellcolor[rgb]{ 1,  1,  1}\textbf{BW} & \cellcolor[rgb]{ 1,  1,  1}\textbf{RTT} & \cellcolor[rgb]{ 1,  1,  1}\textbf{Jitter} & \cellcolor[rgb]{ 1,  1,  1}\textbf{Buffer} & \cellcolor[rgb]{ 1,  1,  1}\textbf{SNR} \\
    RMSE  & 0.004 & 3.4   & 6.5   & 47.4  & 6.7   & 5.1   & RMSE  & 0.04  & 3.7   & 30.9  & 4.5   & 28.2  & 4.79  & RMSE  & 0.01  & 5.6   & 100.9 & 23.6  & 79.8  & 6.8 \\
    MAE   & 0.003 & 1.64  & 4.4   & 35.5  & 5.5   & 3.34  & MAE   & 0.01  & 1.49  & 14.2  & 2.4   & 17.1  & 2.9   & MAE   & 0.006 & 4.4   & 54.5  & 13.3  & 66.3  & 4.4 \\
    MAPE (\%) & \textcolor[rgb]{ 1,  0,  0}{\textit{0.07}} & 3.4   & \textcolor[rgb]{ 1,  0,  0}{\textit{3.5}} & 7.0   & \textcolor[rgb]{ 1,  0,  0}{\textit{12.0}} & 33.0  & MAPE (\%) & \textcolor[rgb]{ 1,  0,  0}{\textit{0.2}} & 25.0  & \textcolor[rgb]{ 1,  0,  0}{\textit{7.0}} & \textcolor[rgb]{ 1,  0,  0}{\textit{3.6}} & 23.0  & 50.0  & MAPE (\%) & \textcolor[rgb]{ 1,  0,  0}{\textit{0.1}} & 14.0  & 20.0  & 13.0  & 30.0  & \textcolor[rgb]{ 1,  0,  0}{\textit{18.0}} \\
    \bottomrule
    \bottomrule
    \end{tabular}%
}
	\label{tab:tabellona}
\end{table*}%

We start by visually analyzing the behavior of the various presented techniques for two specific VoIP flows identified by their specific codecs, namely G.722 and G.729 (for space constraints we omit the visualization for remaining codecs but a summary of performance results for each codec is reported in the Table \ref{tab:tabellona}). The two aforementioned codecs represent two extreme trade-off choices between conversation quality and bandwidth utilization. Indeed, among the codecs used in our experiments, G.722 provides the better audio quality (for instance, in terms of MOS) but the bandwidth consumption is not very efficient (bit rate of $64$ kb/s). In contrast, G.729 offers a slightly lower audio quality but allows a greater bandwidth saving with just $8$ kb/s of bit rate. 

The panels of Figs. \ref{fig:forecast} and \ref{fig:forecast_g729} show the temporal behavior of each variable for codecs G.722 and G.729, respectively. Superimposed onto the actual values of variables (black dashed lines) we report, with different colors as specified in the figures legends, the behavior of each forecasting technique described in the previous section. 
In each panel of Figs. \ref{fig:forecast} and \ref{fig:forecast_g729}, the forecasting zone (the gray area on the right) defines the area where each technique tries to predict future values. 

In order to highlight the behavior of VAR compared to the learning techniques, we also report, in shaded pale red, the $95\%$ forecast intervals (for the VAR) which represent an estimate of the intervals where we expect a future value will fall. Since the interval width amounts to $1.96 \cdot \sigma_\epsilon$ (with $\sigma_\epsilon$ the standard deviation of residuals for each time series, see \cite{lutk}), the shape and the width of each interval strongly depends on the residuals behavior. For instance, as regards the MOS - G.722 codec case (see the first panel in Fig. \ref{fig:forecast}), the low residual standard deviation directly results in a narrow forecast interval. Conversely, the unexpected peak of RTT - G.722 codec case (see the third panel in Fig. \ref{fig:forecast}) at about $415$ seconds has a negative impact on the prediction accuracy (for all examined techniques) and, in turn, implies a growth of residual standard deviation. This directly translates into a broad forecast interval. 
	
The first aspect to highlight is that the actual behavior of variables basically depends on two factors which impact on the performance forecast: the codec type and the network conditions. For instance, the overall bandwidth consumption amounts, on average, to $90$ kb/s in the G.722 case (top-middle panel of Fig. \ref{fig:forecast}), whereas it is around $30$ kb/s in the G.729 case (top-middle panel of Fig. \ref{fig:forecast_g729}). In principle, this implies that more fluctuations are possible in the G.722 case due to a wider span of values. Unfortunately, even if a codec is able to guarantee a kind of temporal ``stability'', the high unpredictability of network conditions is the main responsible of fluctuations which are very challenging to predict due to their extremely time-variant behavior. 

This notwithstanding, in both Figs. \ref{fig:forecast} and \ref{fig:forecast_g729} we can observe that each technique is able to produce a satisfying forecast of the original variables by remaining within the area delimited by the forecast intervals.

By visual inspection, we observe that the VAR (red curve) shows good performance when the time series do not exhibit excessive fluctuations. This notwithstanding, when important fluctuations are present, the VAR model is able to follow the mean value of the oscillating time series (see, for instance, the case of MOS - G.729 codec case).  
The reason is that VAR is governed by a set of linear equations (see (\ref{eq:ar})), thus it can suffer when representing some non-linear behaviors. Occasionally, also the MLP technique shows slight difficulty to fit the original values, once again due to the underlying linear model (see (\ref{eq:neuron})). 
Such a behavior emerges in particular in Fig. \ref{fig:forecast} where the MLP prediction moves a bit away from the corresponding forecast intervals.
Conversely, remaining techniques show enough good adaptation to fluctuations, and, in particular, the deep-based techniques whose internal structure allows to keep the state at time $t-1$ to improve the prediction at time $t$. 

To better quantify the behavior of each technique, we have evaluated the performance for each voice flow (namely for each codec), for each technique, and for each time-based variable as shown in Table \ref{tab:tabellona}. Each sub-table contains the performance per voice flow in terms of test Root Mean Square Error (RMSE), test Mean Absolute value of Errors (MAE), Mean Absolute Percentage Error, defined, respectively, as
\beq
RMSE_j=\sqrt{\frac{\sum_{t=1}^{L} (y_{jt}-\hat{y}_{jt})^2}{L}},
\eeq 

\beq
MAE_j=\frac{\sum_{t=1}^{L}|y_{jt}-\hat{y}_{jt}|}{L},
\eeq
\beq
MAPE_j= \frac{100}{L} \sum_{t=1}^{L}  \bigg |  \frac{y_{jt}-\hat{y}_{jt}}{y_{jt}} \bigg |.
\eeq
Such metrics are computed for each time series $j=1,...,N$, with $N=6$ and $L$ the time series length. These three indicators are often used jointly when evaluating the forecasting accuracy. The RMSE is a quadratic score rule which gives a relatively high weight to large errors since these errors are squared before they are averaged. The MAE is a linear score rule designed to equally weight the individual differences. The MAPE includes a normalization to actual values and is expressed as percentage. Being such an indicator often used as a summarizing metric, to easily pinpoint the best forecasting technique in Table~\ref{tab:tabellona}, the corresponding MAPE value is indicated in red.

For each voice flow we have repeated a lag analysis (just as seen in Sect. \ref{subsect:var}), and we have reported the optimal lag value $p^*$ close to VAR model in each sub-table of Table \ref{tab:tabellona}. 

Let us start to notice some general facts valid for all the experiments. For each voice flow, we can notice that the three performance indicators (RMSE, MAE, and MAPE) exhibit very different ranges for each variable. For instance, in the case of MOS, RMSE and MAE never reach the value $1$. This is due to the fact that all the chosen codecs guarantee a good perceived quality, with a MOS varying within a limited range of values (MOS values never lie below $4$). This directly reflects into low values of all the indicators.

Conversely, RTT varies within a great range of values with some unusual peaks due to the temporary network conditions (see, e.g., the peaks of RTT in green at about $t=30$ s and $t=415$ s in Fig. \ref{fig:train_test_all}). In such cases, all the forecasting techniques are (obviously) not able to predict this behavior, thus the prediction error is quite large. Such a condition directly reflects onto performance indicators and, in particular, onto RMSE whose values are high and hugely different between them since RMSE tends to magnify large errors.  
Yet, the SNR exhibits a quite standard smooth behavior with weak oscillations and not unusual peaks. Thus, the performance indicators are not dramatically high, indicating a satisfying forecasting accuracy. 

Indeed, if we compare the performance accuracy for each technique, we can observe that: VAR technique, as also mentioned before, could return satisfactory result when the time series does not exhibit too many non-linearities; on average, deep techniques such as LSTM, GRU, and CNN produce satisfactory results (see MAPE values in red). For LSTM and GRU the results are justified thanks to the presence of a memory-based internal structure able to keep track of past values. For CNN, results are justified by the presence of a convolutional structure able to derive the most significant temporal features. Remaining deep based techniques (RNN and MLP) exhibit less performing accuracy results due to their naive internal structure which does not exploit any particular characteristic of temporal data. 
Among standard machine learning techniques, XGBoost exhibits the best performance since it relies on a combination of ensemble models useful to improve the quality of prediction.

\begin{figure}[t!]
	\centering
	\captionsetup{justification=centering}
	\includegraphics[scale=0.38]{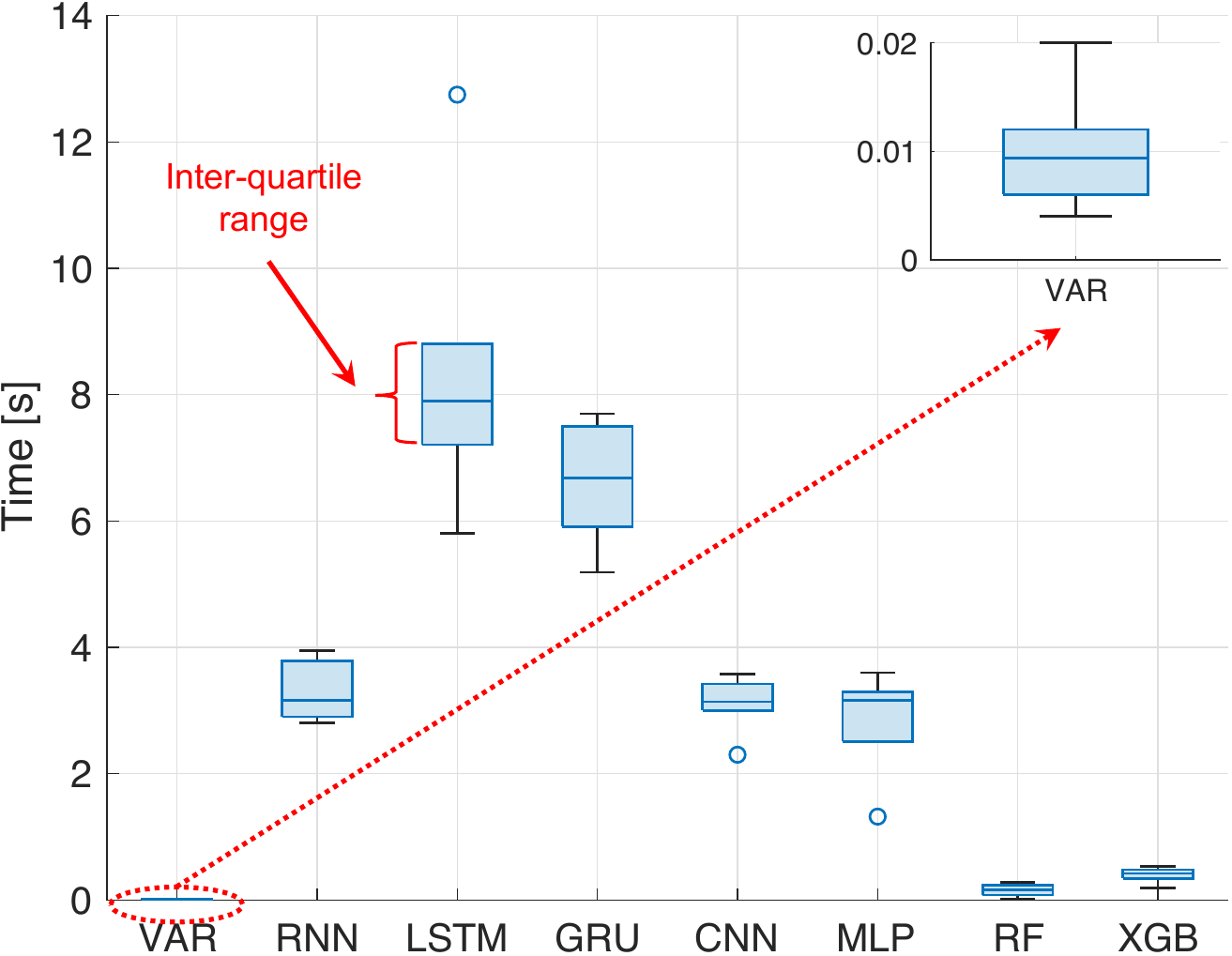}
	\caption{Box-plots of the elapsed times for each forecasting technique applied to all the available voice flows.}
	\label{fig:times}
\end{figure}

The performance analysis should be complemented by a time evaluation to better compare the considered techniques. Figure \ref{fig:times} shows the results of such a comparison obtained by exploiting a platform equipped with $2$ virtual CPUs (Intel Xeon $@$2.30 GHz) and 13 GB of RAM. In this time analysis we take into account all the experiments (namely, all the voice flows) in order to highlight possible variability through a boxplot representation. 
First, such an analysis reveals that the VAR technique is able to perform the forecasting in few milliseconds as shown within the top-right inset. 
This is basically due to the fact that VAR is built through a linear combination of lagged values to forecast the next sample. In contrast, deep methods require more time (in particular during the training time) to perform the forecasting due to their internal structure which can be more or less complicated (e.g., memory-based cells, convolution operations). In the middle we find XGB and RF which, relying on an optimized tree-based structure, are quite fast.
The boxplot representation also highlights that, when using a technique with a complex internal structure (typically, deep-based techniques) the time variability directly increases. This is obviously connected to the fact that a more complex structure may produce higher delays. This behavior is captured through the inter-quartile range (IQR) defined as the difference between third and first quartiles. Large IQR value imply more dispersed values. In case of deep-based techniques we observe the following IQR values: RNN ($0.88$), LSTM ($1.59$), GRU ($1.58$), CNN ($0.42$), MLP ($0.79$). In case of standard learning techniques we have: RF ($0.15$), XGB ($0.14$). Finally, VAR is the more stable having the smallest IQR value amounting to $0.006$. 

\subsection{Main Findings}
Through the proposed assessment we are able to infer some general considerations about the evaluated forecasting techniques. First of all, we can reasonably say that there is not a definitive winner, since the forecasting complexity does not allow to select an outperforming technique in an absolute sense. An insightful comparison can be made between the statistical approach (represented by VAR) and the learning techniques. 

First, we highlight that the VAR method allows a complete control on the analytical structure of each time series. In particular, its ancillary analyses (e.g., residuals, impulse response) provide deep insights about the time series composition and their mutual relationships. In contrast, the data-driven approach adopted by learning techniques does not allow to capture many analytical details. 
For instance, guaranteeing the stationarity condition (not required by learning approaches) allows to obtain useful descriptors (mean, variance, correlation) of the future behavior of a time series. Conversely, in case the stationarity condition is violated (namely if the series is consistently increasing over time) the sample mean and variance will grow with the sample size, and will tend to underestimate the mean and variance in future periods. 

Second, VAR is a good choice in case the time series exhibits a good stability over time, or when the observation time is wider than tens of minutes (e.g., per-month or per-year). In this latter case, in fact, the temporal irregularities tend to be smoother, and the linear combination of past lagged values offer better performance. Conversely, being intrinsically adaptive, learning techniques are more responsive in presence of network parameters fluctuations. 
Furthermore, VAR offers challenging performance in terms of compute times due to the simplicity of the model. 
On the other hand, deep recurrent methods (RNN, LSTM, GRU, CNN, MLP) exhibit slower computation times along with high temporal uncertainty (high IQR values) mainly due to the complex internal structure. Among standard ML techniques, XGBoost offers an interesting trade-off between accuracy and time.

We finally notice that, differently from all the learning techniques, VAR does not need any hyper-parameter tuning (other than the optimal lag) which, if not accurate, could lead to poor performance.

\section{Conclusion}
\label{sec:concl}
In this work we tackle the problem of forecasting mobile VoIP traffic in a real cellular environment. The main purpose is to provide precious information to network operators allowing them to optimize the network planning in mobile environments.
In particular, we characterize the temporal evolution of the most important QoS/QoE descriptors of VoIP traffic through a multivariate time series assessment. Forecasting techniques such as Vector Autoregression and machine learning approaches have been compared to highlight {\em pros} and {\em cons} both in terms of performance and times.


The work presents a series of novelty elements. First, we propose a multivariate time series characterization of network descriptors, an approach currently used by econometricians to model and predict the market stock evolution. Through such an approach it is possible to  analytically capture the interdependencies among the stochastic processes which govern the network variables behavior. Then, the time series problem has been turned into a supervised learning framework through the sliding window technique. 

Such  reframing of the problem is useful to: $i)$ reinterpret the classic concepts of training/test sets in terms of temporal values of a time series aimed at forecasting future values of network descriptors; $ii)$ compare in a critical manner statistical techniques (here represented by the VAR model) and machine learning methods. Results show that VAR is the optimal choice when a complete analytical control on the variables is needed, when the network fluctuations are not so persistent, or when strict elaboration time constraints are present. In contrast, learning-based techniques provide excellent accuracy in case of network instability due their data-driven approach.

Finally, the whole assessment is supported by an experimental campaign in a real mobility LTE-A environment, where through the evolved RTCP-XR protocol, we are able to derive network metrics typically neglected in the literature (e.g., MOS, SNR, playout delay buffer).

Such a work remains open for future investigations along several directions: $i)$ the main techniques adopted for this analysis could be extended to technologies such as $5$G as they become more pervasive and with the possibility of acquiring data from real settings; $ii)$ many derived models could be used as benchmark to design more realistic network simulators; $iii)$ new parameters such as the car's speed could be gathered and related to the behavior of the VoIP metrics; $iv)$ it could be possible to repeat the whole analysis in a transformed domain (e.g., wavelet domain in place of time domain).

\vspace{-30pt}

\begin{IEEEbiography}[{\includegraphics[width=1in,height=1in,clip,keepaspectratio]{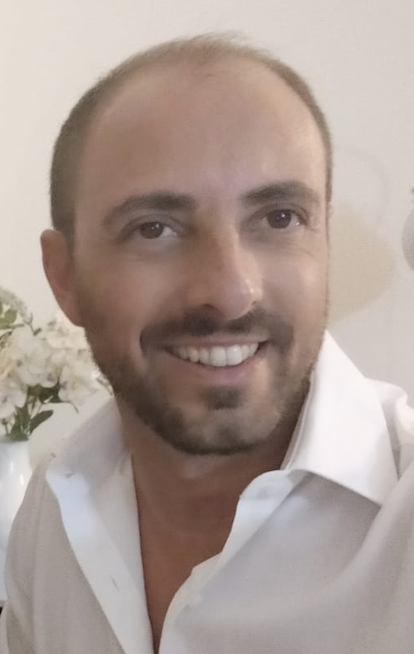}}] 
	{Mario Di Mauro}  (SMIEEE’21) received the Ph.D. degree in information engineering in 2018 from University of Salerno (Italy). From 2007 to 2012 he was with Research Consortium on Telecommunications (formerly Ericsson Lab Italy) as an industrial researcher.
	He is an Assistant Professor in Telecommunications at University of Salerno. His main fields of interest include: network availability and security, data analysis for telecommunication infrastructures.
\end{IEEEbiography}

\vspace{-30pt}

\begin{IEEEbiography}[{\includegraphics[width=1in,height=1in,clip,keepaspectratio,angle=270]{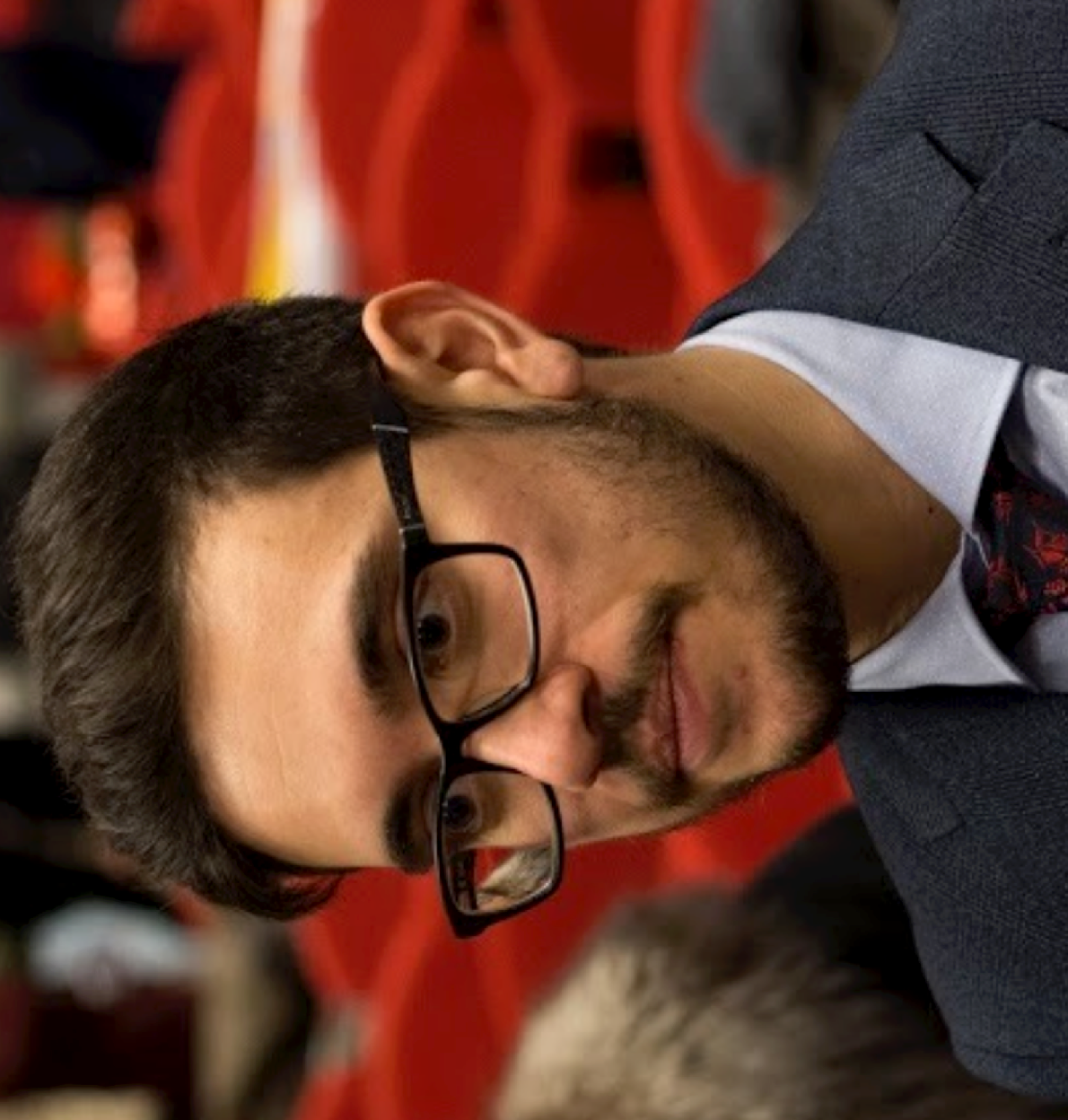}}] 
	{Giovanni Galatro} received the Laurea degree (summa cum laude) in information engineering from the University of Salerno (Italy) in 2018. In 2017 he got a scholarship with Telecommunication and Applied Statistics groups, focused on the analysis of modern telecommunication infrastructures. He is actually a cloud Engineer at IBM.
\end{IEEEbiography}

\vspace{-30pt}

\begin{IEEEbiography}[{\includegraphics[width=1in,height=1in,clip,keepaspectratio]{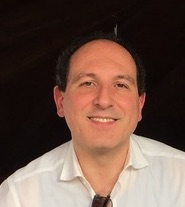}}]
	{Fabio Postiglione} is currently an Associate Professor of Statistics at University of Salerno (Italy). He received his Ph.D. degree in Information Engineering from University of Salerno (Italy) in 2005. His main research interests include degradation analysis, lifetime estimation, reliability and availability evaluation of complex systems (telecommunication networks, fuel cells), Bayesian statistics and data analysis.  
	He has co-authored over 120 papers, mainly published in international journals.
\end{IEEEbiography}

\vspace{-20pt}

\begin{IEEEbiography}[{\includegraphics[width=1in,height=1in,clip,keepaspectratio]{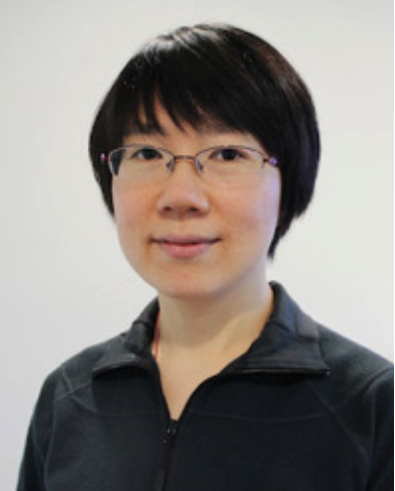}}]
	{Wei Song} received the Ph.D. degree in computer science from Queensland University of Technology (QUT), Brisbane, QLD, Australia, in 2012.
	From 2012 to 2015, she was a Research Fellow with QUT. In 2016, she became a Professor at Shanghai Ocean University, China. She is the Co-Founder of the Joint Intellisensing Lab, Shanghai Ocean University. Her research interests include image/video processing for underwater computer vision, big data analysis, remote sensing image analysis, and its application in marine fields.
	Dr. Song was rewarded as “Eastern Scholar” by Shanghai Institutions of Higher Learning, Shanghai, in 2016.
\end{IEEEbiography}

\vspace{-20pt}

\begin{IEEEbiography}[{\includegraphics[width=1in,height=1in,clip,keepaspectratio]{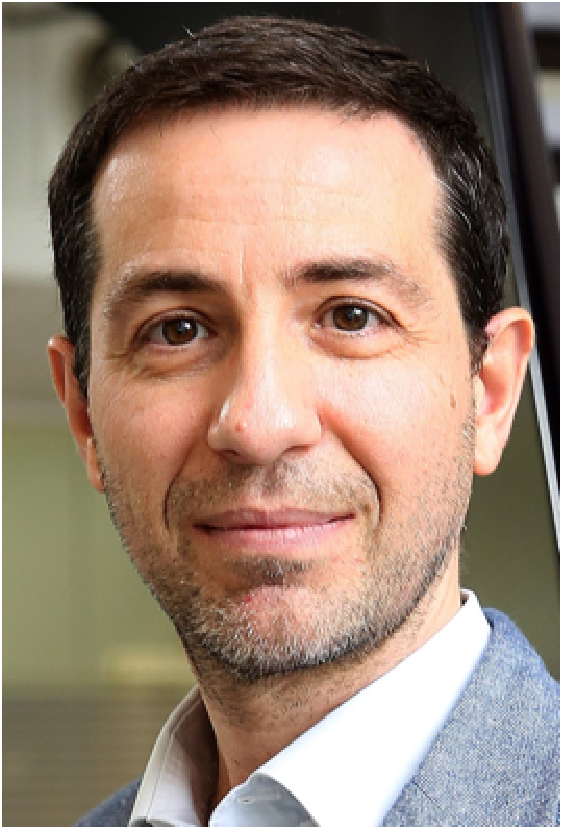}}]  {Antonio Liotta} (SMIEEE’15) is a Full Professor of Data Science at the Free University of Bozen-Bolzano, Italy. His team is at the forefront of influential research in data science and artificial intelligence, specifically in the context of smart cities, IoT, and smart sensing. He is renowned for his contributions to miniaturized machine learning, and artificial neural networks acceleration via network science methods. Antonio has over 350 publications to his credit, and is the Editor-in-Chief of the Springer Internet of Things book series.
\end{IEEEbiography}

\end{document}